\documentclass{ar-1col}

\jname{Xxxx. Xxx. Xxx. Xxx.}
\jvol{AA}
\jyear{YYYY}
\doi{10.1146/((please add article doi))}

\begin{document}

\markboth{Nesvorn\'y}{Early Solar System Dynamics}

\title{Dynamical Evolution of the Early Solar System}

\author{David Nesvorn\'y$^1$
\affil{$^1$Department of Space Studies, Southwest Research Institute, Boulder, USA, CO 80302; email: davidn@boulder.swri.edu}}

\begin{abstract}
Several properties of the Solar System, including the wide radial spacing of the giant planets, can be explained if planets 
radially migrated by exchanging orbital energy and momentum with outer disk planetesimals. Neptune's planetesimal-driven 
migration, in particular, has a strong advocate in the dynamical structure of the Kuiper belt. A dynamical instability is 
thought to have occurred during the early stages with Jupiter having close encounters with a Neptune-class planet. As a result 
of the encounters, Jupiter acquired its current orbital eccentricity and jumped inward by a fraction of an au, as required for 
the survival of the terrestrial planets and from asteroid belt constraints. Planetary encounters also contributed to capture 
of Jupiter Trojans and irregular satellites of the giant planets. Here we discuss the dynamical evolution of the early Solar 
System with an eye to determining how models of planetary migration/instability can be constrained from its present architecture.   
\end{abstract}

\maketitle

\tableofcontents

\section{INTRODUCTION}
The first Solar System solids condensed 4.568 Gyr ago (see Kleine et al. 2009 for a review). This is considered 
as time zero in the Solar System history ($t_0$). Jupiter and Saturn have massive gas envelopes and must have 
formed within the lifetime of the protoplanetary gas disk. From observations we know that the protoplanetary gas 
disks last 2-10 Myr (e.g., Williams \& Cieza 2011). Assuming that the Solar System is typical, 
Jupiter and Saturn should thus have formed within 2-10 Myr after $t_0$ (Kruijer et al. 2017). Geochemical constraints 
and numerical modeling suggest that the terrestrial planet formation ended much later, probably some 30-100 Myr 
after the gas disk dispersal (e.g., Jacobson et al.~2014). 

The subject of this review is the dynamical evolution of planets and small bodies in the Solar System
{\it after} the dispersal of the protoplanetary gas nebula. Here we are therefore not primarily concerned 
with the growth and gas-driven migration of planets (for that, see reviews of Kley \& Nelson 2012 and
Youdin \& Kenyon 2013). The earliest epochs are obviously relevant, because they define the initial conditions 
from which the Solar System evolved. Ideally, this link should be emphasized, but physics of the protoplanetary disk 
stage is not understood well enough to make definitive predictions (except those discussed in Section 3). Instead, 
a common approach to studying the early dynamical evolution of the Solar System is that of reverse engineering, 
where the initial state and subsequent evolution are deduced from various characteristics of the present-day 
Solar System.
\section{PLANETESIMAL-DRIVEN MIGRATION}
Planetary formation is not an ideally efficient process. The growth of planets from smaller disk constituents
can be frustrated, for example, when the orbits become dynamically excited. The accretion of bodies in 
the asteroid belt region (2-4 au) is thought to have ended when Jupiter formed and migrated in the gas disk 
(Walsh et al. 2011). The growth of bodies in the region of the Kuiper belt ($>$30 au), on the other hand, progressed 
at a leisurely pace, because the accretion clock was ticking slowly at large orbital periods, and probably 
terminated, for the most part, when the nebular gas was removed. As a result, in addition to planets, the populations 
of small bodies --commonly known as {\it planetesimals}-- emerged in the early Solar System. 

The gravitational interaction between planets and planetesimals has important consequences. For example, the 
gravitational torques of planets on planetesimals can generate the apsidal density waves in planetesimal disks (e.g., 
Ward \& Hahn 1998). A modest degree of orbital excitation arising in a planetesimal disk at orbital resonances
and/or from gravitational scattering between planetesimals can shut down the wave propagation. In this 
situation, the main coupling between planets and planetesimals arises during their close encounters when they 
exchange orbital momentum and energy. 

Consider a mass $m$ of planetesimals ejected from the Solar System by a planet of mass $M$ and orbital radius 
$r$. From the conservation of the angular momentum it follows that the planet suffers a decrease $\delta r$ 
of orbital radius given by $\delta r/r \simeq -C m/M$, where $C$ is a coefficient of the order of unity 
(Malhotra 1993). For example, if Jupiter at $r\simeq5$ au ejects 15 $M_\oplus$ of planetesimals, where 
$M_\oplus$ is the Earth mass, it should migrate inward by $\delta r \simeq -0.2$ au. Conversely, outer
planetesimals scattered by a planet into the inner Solar System would increase the planet's orbital radius. 

\begin{figure}[t]
\includegraphics[width=5.0in]{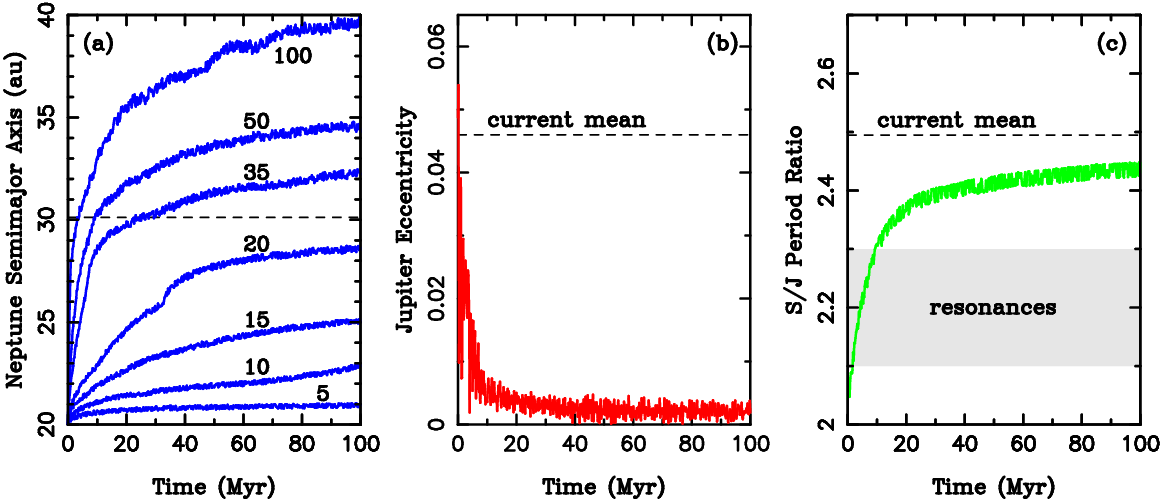}
\caption{Planetesimal-driven migration in the outer Solar System. The outer planets were placed into a compact 
configuration with the semimajor axes $a_5=5.4$ au, $a_6=8.7$ au, $a_7=15.0$ au and $a_8=20.0$ au (index denotes
planets in order of the heliocentric distance). The outer planetesimal disk at 20-30 au was resolved by 10,000 
bodies and was given different masses in different $N$-body integrations (labeled in Earth masses, $M_\oplus$, 
in panel a). Panel (a): Outward migration of Neptune in different cases. The dashed line indicates the current
mean semimajor axis of Neptune ($e_8=30.1$ au). Panel (b): Jupiter's eccentricity becomes quickly damped by dynamical 
friction with scattered planetesimals. Panel (c): The evolution of Saturn/Jupiter period ratio, $P_6/P_5$, for 
$M_{\rm disk}=35$~$M_\oplus$. Secular resonances with the terrestrial planets and asteroids occur in the gray region 
($2.1<P_6/P_5<2.3$).}   
\label{migra}
\end{figure}

Numerical simulations of planetesimal scattering in the outer Solar System show that Neptune, Uranus and Saturn 
tend to preferentially scatter planetesimals inward and therefore radially move outward. Jupiter, on the other hand, 
ejects planetesimals to Solar System escape orbits and migrates inward (Fern\'andez \& Ip 1984). This
process is known as the {\it planetesimal}-driven migration. A direct evidence for the planetesimal-driven 
migration of Neptune is found in the Kuiper belt, where there are large populations of bodies in orbital resonances
with Neptune (e.g., Pluto and Plutinos in the 3:2 resonance with Neptune; Malhotra 1993, 1995; Section 11).

The timescale and radial range of the planetesimal-driven migration depends on the total mass and distribution of
planetesimals (Hahn \& Malhotra 1999). Two migration regimes can be identified (Gomes et al. 2004). If the 
radial density of planetesimals exceeds certain value, the migration is self-sustained and Neptune proceeds to 
migrate outward (upper curves in \textbf{Figure \ref{migra}a}). 
If, on the other hand, the radial mass density of planetesimals is low, the planet runs out of 
fuel and stops (the so-called damped migration; Gomes et al. 2004). The critical density is determined by the 
dynamical lifetime of planetesimals on planet-crossing orbits. 

The critical mass density is near 1-1.5 $M_\oplus$ au$^{-1}$ but depends on other parameters as well. For 
example, a 20 $M_\oplus$ planetesimal disk extending from 20 au to 30 au with a surface density 
$\Sigma \propto 1/r$ is clearly super-critical (radial mass density 2 $M_\oplus$ au$^{-1}$). Neptune's migration 
is self-sustained in this case and Neptune ends up migrating to the outer edge of the disk at 30 au. If the disk mass 
is higher than that, Neptune can even move beyond the original edge of the disk. A 5 $M_\oplus$ planetesimal disk 
with the same parameters, on the other hand, is sub-critical (mass density 0.5 $M_\oplus$ au$^{-1}$) and Neptune's 
migration stalls just beyond 20 au (\textbf{Figure \ref{migra}}). The planetesimal disk survives in the latter case, 
which may be relevant for the long-lived debris disks observed around other stars (Wyatt 2008). 

For super-critical disks, the speed of planetary migration increases with planetesimal mass density. The Kuiper
belt constrains require that the characteristic $e$-folding timescale of Neptune's migration, $\tau$, satisfied 
$\tau \geq 10$ Myr (Section 11). This implies that the planetesimal disk at 20-30 au had mass $M_{\rm disk}\leq20$ 
$M_\oplus$. For Neptune to migrate all the way to 30 au, as discussed above, $M_{\rm disk}\geq15$ $M_\oplus$. Together, 
this implies $M_{\rm disk}\simeq15$-20 $M_\oplus$ for the planetesimal disk between 20-30 au. Beyond 30 au, the 
planetesimal disk must have become sub-critical. The disk may have been truncated, for example, by 
photoevaporation (e.g., Adams 2010; see Section 11).      
\section{PLANETARY INSTABILITY}
Planets form on nearly circular and coplanar orbits. The planetary orbits remain nearly circular and coplanar 
during planetesimal-driven migration, because the collective effect of small disk planetesimals on planets
is to damp any excess motion due to eccentricity or inclination. This so-called dynamical `friction' drives planetary 
orbits toward $e=0$ and $i=0$ ($e_5=0.002$ and $e_6=0.005$ at the end of the simulation with $M_{\rm disk}=35$ $M_\oplus$; 
\textbf{Figure \ref{migra}b}). In contrast, Jupiter and Saturn have current mean eccentricities $e_5=0.046$ and 
$e_6=0.054$, respectively. Also, the orbits of Saturn and Uranus are significantly inclined ($i_6=0.90^\circ$ and 
$i_7=1.0^\circ$). This shows need for some excitation mechanism.

\begin{figure}[t]
\includegraphics[width=4.0in]{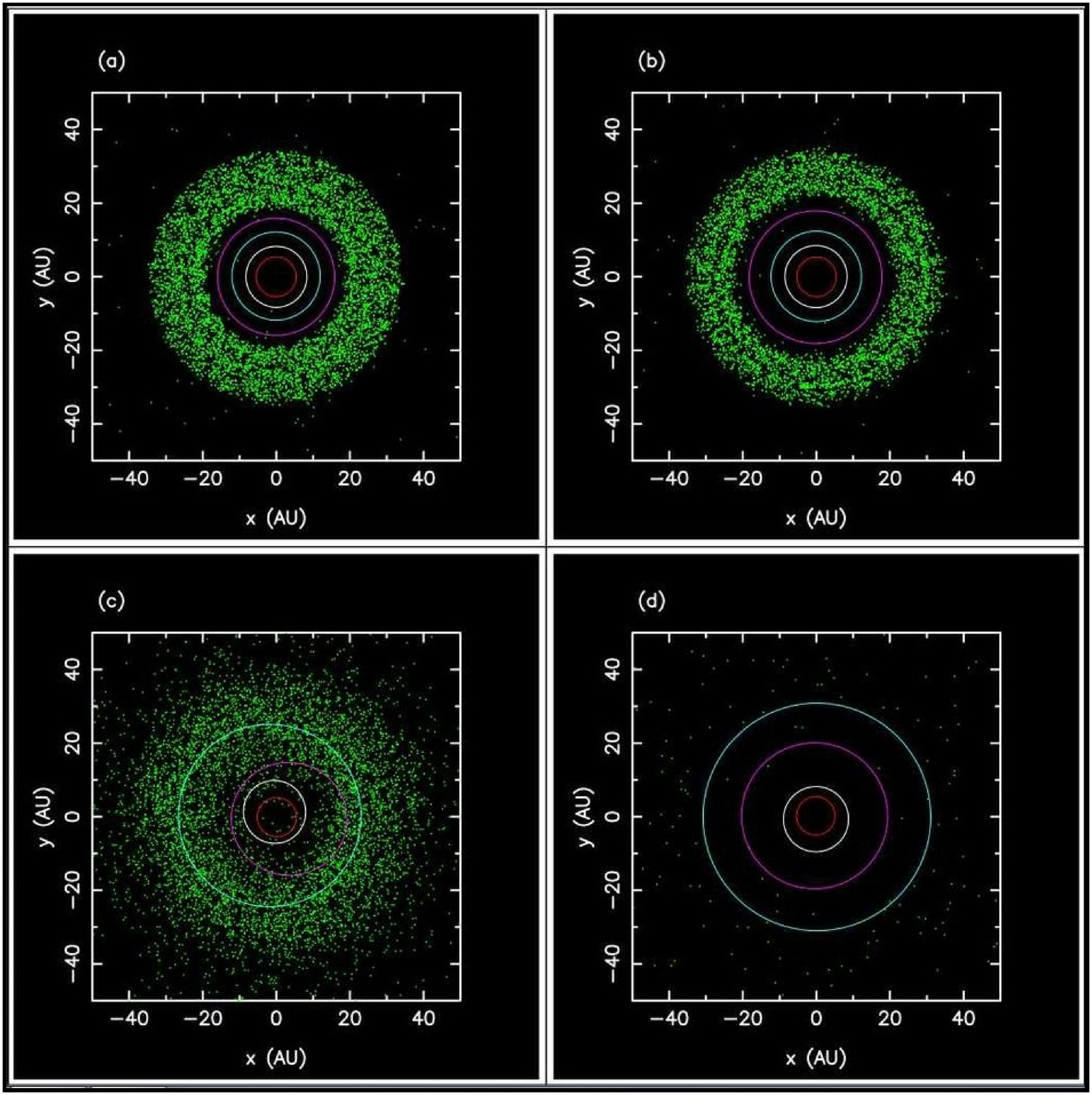}
\caption{The original Nice model simulation showing the outer planets and planetesimals: (a) initial 
configuration; (b) early configuration, before Jupiter and Saturn reach the 2:1 resonance; (c) scattering 
of planetesimals after the orbital shift of Neptune (dark blue) and Uranus (light blue); (d) after ejection 
of planetesimals by planets.}   
\label{nice}
\end{figure}

Tsiganis et al. (2005; also see Thommes et al. 1999) proposed that the excitation of planetary orbits occurred 
when Jupiter and Saturn crossed the 2:1 resonance during the planetesimal-driven migration. Because the planetary 
orbits diverge from each other (Jupiter moves in and Saturn out), the resonance is approached from a direction 
from which capture cannot occur. Instead, the orbits cross the 2:1 resonance and acquire modest eccentricities. This, 
in itself, would not be enough to provide the needed excitation (e.g., the orbits of Uranus and Neptune remained 
unaffected). The 2:1 resonant crossing, however, can trigger a dynamical instability in the outer Solar System 
with Uranus and/or Neptune eventually evolving onto Saturn-crossing orbits. A dynamical excitation of 
orbits then presumably happened during scattering encounters between planets. As a result of planetary encounters, 
Uranus and Neptune were thrown out into the outer planetesimal disk, where they stabilized and migrated to 
their current orbits (\textbf{Figure \ref{nice}}).

This instability model, known as the `original' Nice model (after a city in southern France where the model 
was conceived), has been successful in reproducing the orbits of the outer Solar System planets. It has 
also provided a scientific justification for the possibility that the Late Heavy Bombardment of the Moon 
was a spike in the impact record (Gomes et al. 2005, Levison et al. 2011; Section 13), and offered a convenient 
framework for understanding various other properties of the Solar System (e.g., Morbidelli et al. 2005, 
Nesvorn\'y et al. 2007, Levison et al. 2008; Sections 6-12). 

The evolution of planets during the dynamical instability is stochastic. This means that small changes of the 
initial conditions can lead to different results. It is therefore insufficient to perform one or a few 
dynamical simulations. Instead, a {\it statistical} model must be developed, where many realizations of the 
same initial conditions are tested.

The initial conditions of a model can be informed from the gas-driven migration of planets during the 
protoplanetary disk phase. Hydrodynamic studies show that Jupiter and Saturn underwent a convergent migration 
in the gas disk with their orbits approaching each other. The orbits were subsequently captured into an orbital 
resonance. Under standard conditions (the gas surface density $\Sigma\simeq1700$ g cm$^{-2}$ at 1 au from 
the Minimum Mass Solar Nebula model, MMSN, Weidenschilling 1977, Hayashi 1981; $\alpha$ viscosity $\simeq10^{-3}$-$10^{-2}$, 
Shakura \& Sunyaev 1973; aspect ratio $H/r \simeq 0.05$), the orbits cross the 2:1 resonance without being 
captured, because the convergent migration is too fast for capture to happen. They are eventually trapped in 
the 3:2 resonance (Masset \& Snellgrove 2001, Morbidelli \& Crida 2007, Pierens \& Nelson 2008). 

The 3:2 resonance configuration of the Jupiter-Saturn pair is the essential ingredient of the Grand Tack (GT) 
model (Walsh et al. 2011). In the GT model, Jupiter migrated down to $r\simeq1.5$ au when the 3:2 resonance was 
established. After that, Jupiter and Saturn opened a common gap in the disk, the migration torques reversed their usual 
direction, and Jupiter, after executing the sailing maneuver of tacking, moved outward to beyond 5~au. 
The GT model helps to explain the small mass of Mars and asteroid belt, and mixing of the taxonomic types in 
the asteroid belt (Gradie \& Tedesco 1982). Subsequent studies showed that capture of Jupiter and Saturn in the 
2:1 resonance would be possible if planets migrated slowly in a low-mass and low-viscosity disk (Pierens et al. 2014). 
It is harder in this case, however, to obtain the torque reversal and stable outward migration.

Given these results, it is reasonable to expect that Jupiter and Saturn emerged from the protoplanetary disk 
with orbits locked in the 3:2 resonance (or, somewhat less likely, in the 2:1 resonance). The initial orbits of Uranus 
and Neptune should have been resonant as well (Morbidelli et al. 2007). In a MMSN disk with modest viscosity, capture
of ice giants in the 3:2, 4:3 and 5:4 resonances is preferred. In a low-mass disk, instead, the 2:1 and 3:2 resonances 
are preferred. The latter case may apply if Uranus/Neptune formed late, near the end of the protoplanetary disk phase, 
as indicated by their low-mass gas envelopes.

The initial orbits of the outer planets in a fully resonant chain have not been considered in the original Nice 
model. Morbidelli et al. (2007) performed several simulations starting from the initially resonant conditions. 
They showed that the subsequent dynamical evolution of planets is qualitatively similar to that reported in 
Tsiganis et al. (2005). Still, none of these instability/migration models properly accounted for many Solar 
System constraints, including the secular architecture of the outer planet system, survival of the terrestrial 
planets, and orbital structure of the asteroid belt.
\section{JUMPING JUPITER}
In the original Nice model, Jupiter does not participate in the dynamical instability (i.e., there are no 
encounters of Jupiter with other planets). This is desirable because if Saturn had close 
encounters with Jupiter, the mutual scattering between Jupiter and Saturn would lead to a very strong 
orbital excitation, which has to be avoided. The strong dynamical instabilities between massive planets 
are thought to be responsible for the broad eccentricity distribution of the Jupiter-class exoplanets (Rasio 
\& Ford 1996). 

In Tsiganis et al. (2005), the orbital eccentricity of Jupiter, $e_5$, is generated when Jupiter and Saturn cross 
the 2:1 resonance during their planetesimal-driven migration. Additional changes of $e_5$ occur during encounters 
between Saturn and Uranus/Neptune, because the evolution of eccentricities is coupled via the 
Laplace-Lagrange equations (Murray \& Dermott 1999). The Laplace-Lagrange equations applied to 
Jupiter and Saturn yield 
\begin{eqnarray}
e_5 \exp \iota \varpi_5 & = & e_{55} \exp \iota (g_5 t + \phi_5) + e_{56} \exp \iota (g_6 t + \phi_6) \nonumber \\ 
e_6 \exp \iota \varpi_6 & = & e_{65} \exp \iota (g_5 t + \phi_5) + e_{66} \exp \iota (g_6 t + \phi_6) \ ,
\end{eqnarray}
where $e_{ij}$, $g_i$ and $\phi_i$ are amplitudes, frequencies and phases. Specifically, $g_5=4.24$ arcsec 
yr$^{-1}$, $g_6=28.22$ arcsec yr$^{-1}$, $e_{55}=0.044$, $e_{56}=-0.015$, $e_{66}=0.033$ and $e_{65}=0.048$ 
in the present Solar System.

Since $e_{55}>|e_{56}|$, Jupiter's proper eccentricity mode is excited more that the forced mode. 
Conversely, during the 2:1 resonance crossing and Saturn's encounters, the proper mode would end up being 
less excited than the forced one, leaving $e_{55}<|e_{56}|$ (\'Cuk 2007). Therefore, while the excitation of 
Jupiter's eccentricity is adequate in the original Nice model, the partition of $e_5$ in the modal amplitudes 
$e_{55}$ and $e_{56}$ is not. The most straightforward way to excite $e_{55}$ is to postulate that Jupiter have 
participated in planetary encounters with an ice giant, for example, with Uranus (Morbidelli et al. 2009a). 

The slow migration of Jupiter and Saturn past the 2:1 resonance, which is a defining feature of the original 
Nice model (\textbf{Figure \ref{migra}c}), is difficult to reconcile with several Solar System constrains, 
including the low Angular Momentum Deficit (AMD) of the terrestrial planets and the orbital structure of the asteroid belt 
(Sections 7 and 8). These constraints imply that the orbital period ratio of Jupiter and Saturn, $P_6/P_5$, must have discontinuously 
changed from $<$2.1 to at least $>$2.3 (Brasser et al. 2009, Morbidelli et al. 2010), perhaps 
because Jupiter and Saturn had encounters with an ice giant and their semimajor axes changed by a fraction of 
an au. The terrestrial planet constraint could be bypassed if the migration/instability of the outer planets happened 
early (Section 13), when the terrestrial planet formation was not completed, but the asteroid constraint applies 
independently of that. 

These are the basic reasons behind the jumping-Jupiter model. Jupiter's jump can be accomplished 
if Jupiter had close encounters with an ice giant with the mass similar to Uranus or Neptune. The scattering 
encounters with an ice giant would also help to excite the $e_{55}$ mode in Jupiter's orbit, 
as needed to explain its present value. In addition, Jupiter's planetary encounters provide the right framework 
for capture of Jupiter Trojans (Section 9) and irregular satellites (Section 10). The jumping-Jupiter model 
is therefore a compelling paradigm for the early evolution of the Solar System.\footnote{Alternatives, such as a 
the planetesimal-driven migration model of Malhotra \& Hahn (1999), face several problems, including: (1) $e_5$ 
and $e_6$ are not excited enough in these models, (2) the secular resonances sweep over the asteroid 
belt and produce excessive excitation of asteroid inclinations (Morbidelli et al. 2010), and (3) the AMD 
of the terrestrial planet system ends up to be too high (e.g., Agnor \& Lin 2012).}     

If Jupiter and Saturn started in the 3:2 resonance, where $P_6/P_5 \simeq 1.5$, the easiest way to satisfy 
constraints is to have a few large jumps during planetary encounters such that $P_6/P_5$ changed from 
$\simeq$1.5 directly to $>$2.3. Dynamical models, in which the planetesimal-driven migration extracts Jupiter 
and Saturn from the 3:2 resonance and moves $P_6/P_5$ closer to 2 before the instability happens, are 
statistically unlikely (because there is a tendency for the instability to develop early, or not at all). 
If Jupiter and Saturn started in the 2:1 resonance instead, where  $P_6/P_5 \simeq 2$, the required jump 
is smaller and can be easier to accomplish in a numerical model (Pierens et al. 2014). This is the main 
advantage of the 2:1 resonance configuration over the 3:2 resonance configuration. All other considerations
favor 3:2. 
\section{FIVE PLANET MODEL}
The results of dynamical simulations, when contrasted with the observed properties of the present-day Solar System, 
can be used to backtrack the initial conditions from which the Solar System evolved. In one of the most complete 
numerical surveys conducted so far, Nesvorn\'y \& Morbidelli (2012; hereafter NM12) performed nearly $10^4$ simulations 
of the planetary migration/instability starting from hundreds of different initial conditions. A special attention 
was given to the cases with Jupiter and Saturn initially in the 3:2 and 2:1 orbital resonances. They experimented with 
different radial profiles and orbital distributions of disk planetesimals, different disk masses, etc. 

The cases with four, five and six outer planets were tested, where the additional planets were placed onto resonant 
orbits between Saturn and Uranus, or beyond the initial orbit of Neptune. The cases with additional planets 
were considered in NM12, because it was found that they produce much better results than the four-planet case (see 
below).\footnote{The existing planet formation theories do not have the predictive power to tell us how many ice 
giants formed in the Solar System, with some suggesting that as many as five ice giants have formed (Ford \& Chiang 
2007, Izidoro et al. 2015).} The masses of additional planets were set between 0.3 $M_7$ and 3 $M_8$, where $M_7=14.5$ 
$M_{\oplus}$ and $M_8=17.2$ $M_{\oplus}$ are the masses of Uranus and Neptune. 

\begin{figure}[t]
\includegraphics[width=4.5in]{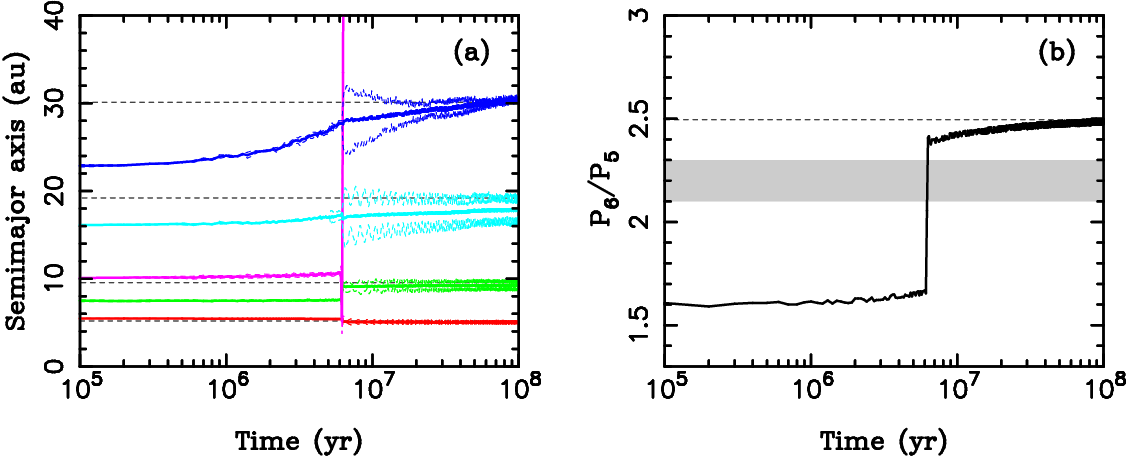}
\caption{Orbital histories of the giant planets from NM12. Five planets were started in the 
(3:2,4:3,2:1,3:2) resonant chain, and $M_{\rm disk}=20$ $M_\oplus$. (a) The semimajor axes (solid lines), 
and perihelion and aphelion distances (dashed lines) of each planet's orbit.  The horizontal dashed lines 
show the semimajor axes of planets in the present Solar System. The final orbits obtained in the model, 
including $e_{\rm 55}$, are a good match to those in the present Solar System. (b) The period ratio $P_6/P_5$. The 
dashed line shows $P_6/P_5=2.49$ corresponding to the present orbits of Jupiter and Saturn. The shaded area 
approximately denotes the zone, where the secular resonances with the terrestrial planets and asteroids 
occur. These resonances are not activated, because the period ratio `jumps' over the shaded area as Jupiter 
and Saturn scatter off of the ejected ice giant.}   
\label{nice2}
\end{figure}

NM12 defined four criteria to measure the overall success of their simulations. First of all, the final 
planetary system must have four giant planets (criterion A) with orbits that resemble the present ones 
(criterion B). Note that A means that one and two planets must be ejected in the five- and six-planet 
planet cases, while all four planets must survive in the four-planet case. As for B, success was claimed 
if the final semimajor axis of each planet was within 20\% to its present value, and the final 
eccentricities and inclinations were no larger than 0.11 and 2$^\circ$, respectively. These thresholds 
were obtained by doubling the current mean eccentricity of Saturn ($e_6=0.054$) and mean inclination 
of Uranus ($i_7=1.02^\circ$). NM12 also required that $e_{55}>0.022$, i.e., at least half of its current 
value~(criterion~C; see discussion in Section 4). Moreover, the $P_{\rm 6}/P_{\rm 5}$ ratio was required to 
evolve from $<$2.1 to $>$2.3 in $\ll$ 1 Myr (Criterion D), as needed to satisfy the terrestrial planet and 
asteroid belt constraints (Sections 7 and 8). The terrestrial planets and asteroids were not explicitly 
included in NM12 to speed up the calculations. 

\begin{figure}[t]
\includegraphics[width=4.5in]{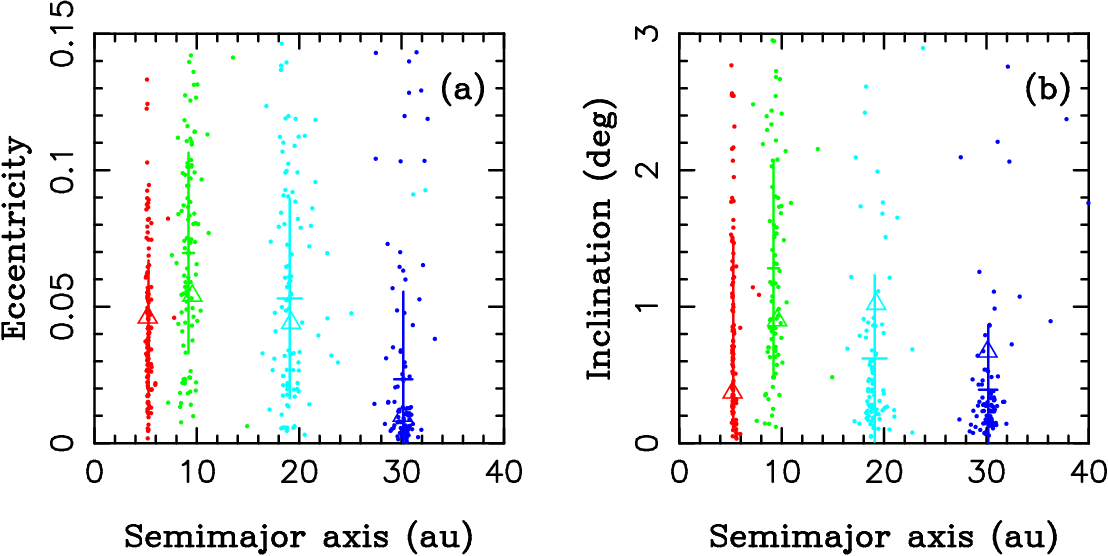}
\caption{Final planetary orbits obtained in 500 simulations with five outer planets started in the (3:2,3:2,2:1,3:2) 
resonant chain and $M_{\rm disk}=20$ M$_\oplus$. The mean orbital elements were obtained by averaging the osculating 
orbital elements over the last 10 Myr of the simulation. Only the systems ending with four outer planets are 
plotted here (dots). The bars show the mean and standard deviation of the model distribution of orbital elements. 
The mean orbits of real planets are shown by triangles. Colors red, green, turquoise and blue correspond to Jupiter, 
Saturn, Uranus and Neptune.}   
\label{disk20}
\end{figure}
   
\textbf{Figure \ref{nice2}} shows an example of a successful simulation that satisfied all four 
criteria. This is a classical example of the jumping-Jupiter model. The instability happened in this 
case about 6 Myr after the start of the simulation. Before the instability, the three ice giants slowly
migrated by scattering planetesimals. The instability was triggered 
when the inner ice giant crossed an orbital resonance with Saturn and its eccentricity was pumped up. Following  
that, the ice giant had encounters with all other planets, and was ejected from the Solar System 
by Jupiter. Jupiter was scattered inward and Saturn outward during the encounters, with $P_6/P_5$ moving 
from $\simeq$1.7 to $\simeq$2.4 in less that $10^5$ yr (\textbf{Figure \ref{nice2}b}). The orbits of Uranus and 
Neptune became excited as well, with Neptune reaching $e_8\simeq0.15$ just after the instability 
($e_8\simeq0.05$-0.15 in all successful models from NM12). The orbital eccentricities were subsequently 
damped by dynamical friction from the planetesimal disk. Uranus and Neptune, propelled by the 
planetesimal-driven migration, reached their current orbits some 100 Myr after the instability. The final 
eccentricities of Jupiter and Saturn were $e_5=0.031$ (modal amplitude $e_{55}=0.030$) and $e_6=0.058$.
For comparison, the mean eccentricities of the real planets are $e_5=0.046$ and $e_6=0.054$. 

NM12 found (see also Nesvorn\'y 2011, Batygin et al. 2012, Deienno et al. 2017) that the dynamical evolution 
is typically too violent, if planets start in a compact resonant configuration, leading to ejection of at least one ice 
giant from the Solar System. Planet ejection could be avoided for large masses of the outer planetesimal 
disk ($M_{\rm disk} > 50$ $M_\oplus$), but a massive disk would lead to excessive dynamical damping 
(\textbf{Figure \ref{migra}b}) and migration regime that violates various constraints (e.g., Section 
11). The dynamical simulations starting with a resonant system of four giant planets thus have a very low 
success rate. In fact, NM12 have not found any case that would satisfy all four criteria in nearly 3000 
simulations of the four planet case. Thus, either the Solar System followed an unusual evolution path
($<$1/3000 probability to satisfy criteria A-D), some constraints are misunderstood, or there were 
originally more than four planets in the outer Solar System. 

Better results were obtained in NM12 when the Solar System was assumed to have five giant planets initially 
and one ice giant, with the mass comparable to that of Uranus and Neptune, was ejected into interstellar space 
by Jupiter (\textbf{Figure \ref{nice2}}). The best results were obtained when the ejected planet was placed into 
the external 3:2 or 4:3 resonance with Saturn and $M_{\rm disk} \simeq 15$-20 M$_\oplus$. The range of possible 
outcomes is rather broad in this case (\textbf{Figure \ref{disk20}}), indicating that the present Solar System 
is neither a typical nor expected result for a given initial state, and occurs, in best cases, with a $\simeq$5\% 
probability (as defined by the NM12 success criteria). [If it is assumed that each of the four NM12 criteria is 
satisfied in 50\% of cases, and the success statistics are uncorrelated, the expectation is $0.5^4=0.063$, or 6.3\%.] 


\begin{summary}[SUMMARY]
In summary of Sections 2-5, the planetesimal-driven migration explains how Uranus and Neptune evolved from 
their initially more compact orbits. Neptune's migration, in particular, is badly needed to understand the 
orbital structure of the Kuiper belt, where orbital resonances with Neptune are heavily populated (Section 11). 
The planetesimal-driven migration, when applied to Jupiter and Saturn, leads to an impasse, because it does 
not explain why Jupiter's present eccentricity (and specifically the $e_{55}$ mode) is significant. The planetesimal-driven 
migration of Jupiter and Saturn also generates incorrect expectations for the terrestrial planet AMD and 
the orbital structure of the asteroid belt. 

The dynamical instability in the outer Solar System, followed by encounters of an ice giant with all other 
outer planets, offers an elegant solution to these problems. The scattering encounters excite 
the orbital eccentricities and inclinations of the outer planets (including the $e_{55}$ mode). 
As a result of the scattering encounters, Jupiter jumps inward and Saturn outward. The inner Solar 
System constraints are not violated in this case. The jumping-Jupiter model is the most easily accomplished if there 
initially was a third ice giant planet, with mass comparable to that of Uranus or Neptune, on a resonant orbit 
between Saturn and Uranus. The orbit of the hypothesized third ice giant was destabilized during the instability 
and the planet was subsequently ejected into interstellar space by Jupiter.
\end{summary}
\section{GIANT PLANET OBLIQUITIES}
The obliquity, $\theta$, is the angle between the spin axis of an object and the normal to its orbital plane.
Here we consider the obliquities of Jupiter and Saturn.\footnote{The terrestrial planets acquired their 
obliquities by stochastic collisions during their formation and subsequent chaotic evolution. 
The obliquities of Uranus and Neptune also do not represent a fundamental constraint on the dynamical evolution
of the early Solar System, because their spin precession rates are much slower than any secular eigenfrequencies of 
orbits. Giant impacts have been invoked to explain the large obliquity of Uranus (e.g., Morbidelli et al. 2012a).} 
The core accretion theory applied to Jupiter and Saturn implies that their primordial obliquities should be small. 
This is because the angular momentum of the rotation of these planets is contained almost entirely in their massive 
hydrogen and helium envelopes. The stochastic accretion of solid cores should therefore be irrelevant for their 
current obliquity values, and a symmetric inflow of gas on forming planets should 
lead to $\theta=0$. The present obliquity of Jupiter is $\theta_5=3.1^\circ$, which is small enough to be roughly 
consistent with these expectations, but that of Saturn is $\theta_6=26.7^\circ$, which is not.

It has been noted (Ward \& Hamilton 2004, Hamilton \& Ward 2004) that the precession frequency of Saturn's spin axis, 
$p_6=-\alpha_6 \cos \theta_6$, where $\alpha_6$ is Saturn's precessional constant (a function of the quadrupole 
gravitational moment, etc.), has a value close to $s_8=-0.692$ arcsec yr$^{-1}$, where $s_8$ is the 8th nodal 
eigenfrequency of the planetary system (Section~7). Similarly, Ward \& Canup (2006) pointed out that $p_5=-\alpha_5 
\cos \theta_5 \simeq s_7$, where $\alpha_5$ is Jupiter's precessional constant, and $s_7=-2.985$~arcsec~yr$^{-1}$. 

While it is not clear whether the spin states of Jupiter and Saturn are actually in the spin-orbit resonances 
at the present (e.g., the current best estimate for Saturn is $|p_6|=0.75$ arcsec yr$^{-1}$, Helled et al. 2009), the 
similarity of frequencies is important, because the spin-orbit resonances can excite $\theta$. This works as follows. 
There are several reasons to believe that $\alpha_6/s_8$ has not remained constant since Saturn's formation. For 
example, it has been suggested that $|\alpha_6/s_8| < 1$ initially, and then evolved to $|\alpha_6/s_8| = 1$, 
when $\alpha_6$ increased as a result of Saturn's cooling and contraction, or because $s_8$ decreased during the 
depletion of the primordial Kuiper belt. If so, the present obliquity of Saturn could be explained by capture of 
Saturn's spin vector in the $p_6 = s_8$ resonance, because the resonant dynamics can compensate for the slow 
evolution of $\alpha_6/s_8$ by boosting $\theta_6$ (Ward \& Hamilton 2004).

While changes of $\alpha_6/s_8$ during the earliest epochs could have been important, it seems more likely that 
capture in the spin-orbit resonance occurred later, probably as a result of planetary migration. 
This is because both $s_7$ and $s_8$ significantly change during the planetary migration and dispersal of 
the outer disk. Therefore, if the spin-orbit resonances had been established earlier, they would not survive to the 
present time. Bou\'e et al. (2009) studied various models for tilting Saturn's spin axis during planetary migration 
and found that the present obliquity of Saturn can be explained by resonant capture if the characteristic migration 
time scale was long and/or Neptune reached high orbital inclination during the instability. 

\begin{figure}[t]
\centering
\begin{minipage}{2.2in}
\centering
\includegraphics[width=2.2in]{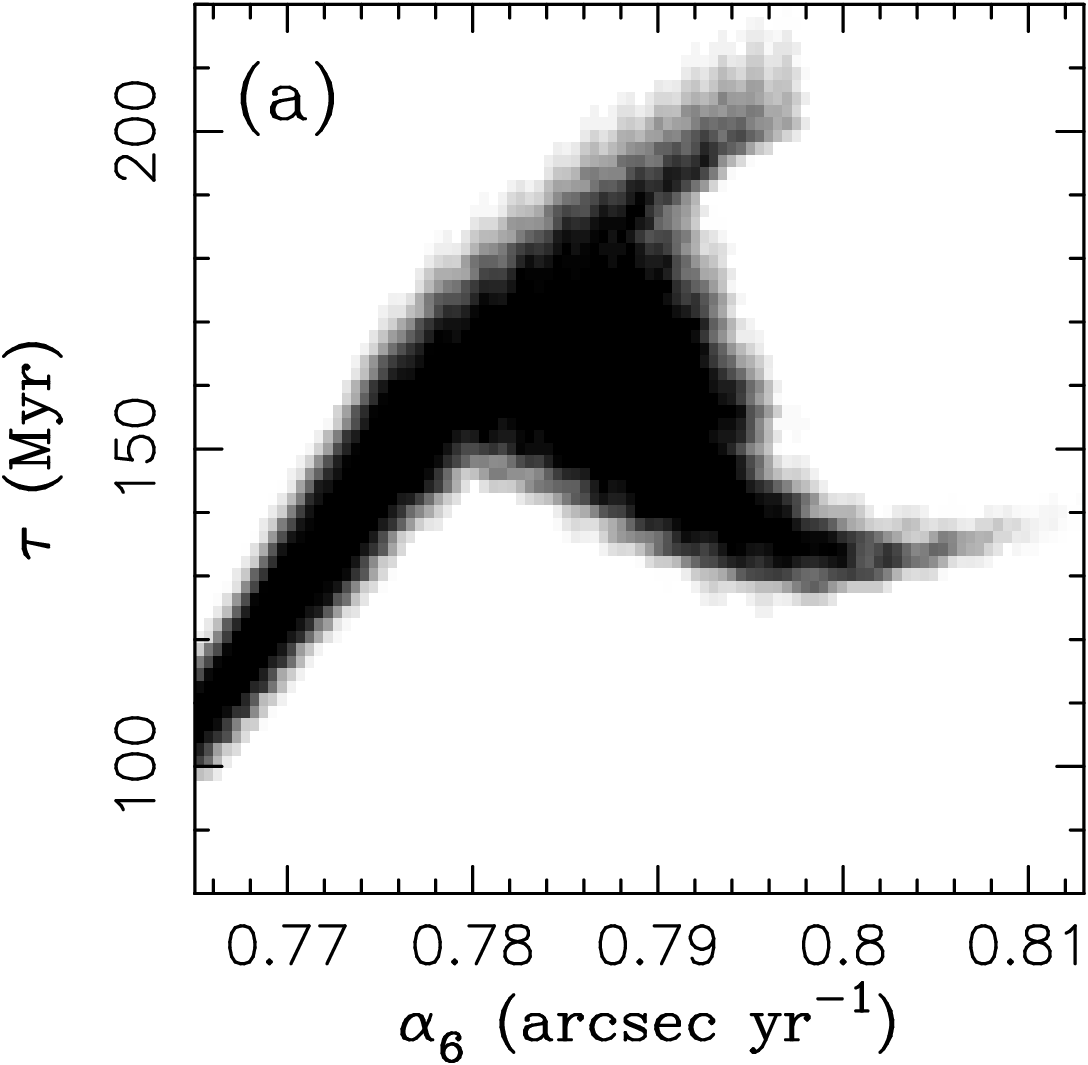}
\end{minipage}
\hspace*{0.1in}
\begin{minipage}{2.2in}
\centering
\vspace*{-4.mm}
\includegraphics[width=2.2in]{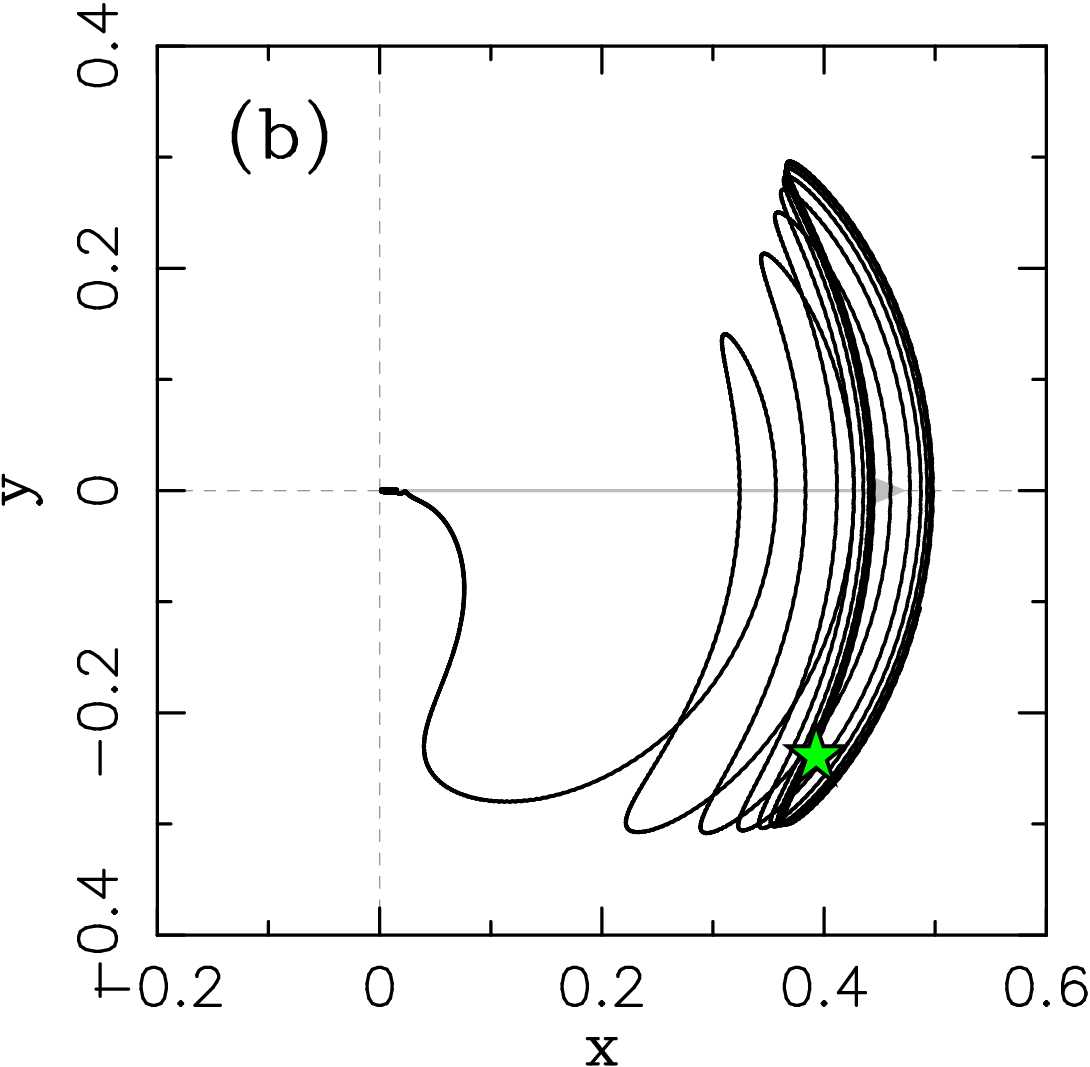}
\end{minipage}
\caption{Left: Saturn obliquity constrain on the planetary migration timescale $\tau$ and precession constant 
$\alpha_6$. Different $\tau$ and $\alpha_6$ values were tested to see how they affect the excitation 
of Saturn's obliquity in the $p_6 = s_8$ spin-orbit resonance. The dark region highlights the parameter 
values which resulted in $\theta_6 \simeq 27^\circ$ and simultaneously provided a sufficiently large 
libration amplitude ($>$$30^\circ$) in the resonance to explain the orientation of Saturn's spin vector. 
Right: Example of Saturn's obliquity excitation for $\alpha_6=0.785$ arcsec yr$^{-1}$ and $\tau=150$ Myr. 
The spin axis $\mathbf{s}$ is projected onto 
$(x,y)$ plane, where $x=\sin \theta \cos \varphi$ and $y=\sin \theta \sin \varphi$, and $\varphi$ defines 
the azimuthal orientation of $\mathbf{s}$. The arrow shows the evolution of the Cassini state $C_2$ over 
the whole length of the simulation (1 Gyr). The present orientation of Saturn's spin vector in the 
$(x,y)$ plane is denoted by the green star. Figures from VN15.} 
\label{obliq}
\end{figure}

In fact, the obliquities of Jupiter and Saturn represent a stronger constraint on the instability/migration
models than was realized before. This is because they must be satisfied {\it simultaneously} (Brasser \& Lee 2015). 
For example, in the initial 
compact configurations of the Nice model, the $s_8$ frequency is much faster than both $\alpha_5$ and $\alpha_6$. 
As $\alpha_5 > \alpha_6$, this means that $s_8$ should first cross $\alpha_5$ to reach $\alpha_6$ during the subsequent 
evolution. This leads to a conundrum, because if the crossing were slow, $\theta_5$ would increase as a result of 
capture into the spin-orbit resonance with $s_8$. If, on the other hand, the evolution were fast, the conditions 
for capture of $\theta_6$ into the spin-orbit resonance with $s_8$ would not be met (Bou\'e et al. 2009), 
and $\theta_6$ would stay small.

A potential solution of this problem is to invoke fast evolution of $s_8$ at the $s_8 = -\alpha_5$ crossing,
and slow evolution of $s_8$ at the $s_8 = -\alpha_6$ crossing. This can be achieved, for example, if the migration 
of outer planets was relatively fast initially, and slowed down later, as planets converged to their current 
orbits. Vokrouhlick\'y \& Nesvorn\'y (2015; hereafter VN15) documented this possibility in the jumping-Jupiter models 
developed in NM12. Recall from Section 5 that the most successful NM12 models feature two-stage migration histories 
with $\tau_1 \sim 10$ Myr before the instability and $\tau_2 \sim 30$-50 Myr after the instability. Moreover, 
the migration tends to slow down relative to a simple exponential at very late stages (effective $\tau \geq 100$ Myr). 

VN15 found that Saturn's obliquity can indeed be excited by capture in the $p_6 = s_8$ spin-orbit resonance 
(Ward \& Hamilton 2004, Hamilton \& Ward 2004, Bou\'e et al. 2009) during the late stages of planetary migration. 
To reproduce the current orientation of Saturn's spin vector, however, specific conditions must be met 
(\textbf{Figure \ref{obliq}}). First, Neptune's late-stage migration must be slow with $100<\tau<200$ Myr. Fast 
migration rates with $\tau<100$ Myr do not work because the resonant capture and excitation of $\theta_6$ do not 
happen.\footnote{Recall that $i_8$ stays below 1$^\circ$ in NM12 such that the high-inclination regime studied 
in Bou\'e et al. (2009) probably does not apply.} Second, for Saturn to remain in the $p_6 = s_8$ resonance today, $\alpha_6<0.8$ arcsec 
yr$^{-1}$, which is lower than the estimate derived from modeling of Saturn's interior ($\alpha_6=0.845$ 
arcsec yr$^{-1}$; Helled et al. 2009). Interestingly, direct measurements of the mean precession rate 
of Saturn's spin axis suggest $\alpha_6=0.81\pm0.05$ arcsec yr$^{-1}$ (see discussion in VN15), which would 
allow for $\alpha_6<0.8$ arcsec yr$^{-1}$ within the quoted uncertainty.

As for Jupiter, the $p_5 = s_8$ resonance occurs during the first migration stage of NM12. To avoid 
resonant capture and excessive excitation of $\theta_5$, either ${\rm d}s_8/{\rm d}t$ must be large 
or $i_{58}$ must be small, where $i_{58}$ is the amplitude of the $s_8$ term in the Fourier expansion
of Jupiter's orbital precession. There are good reasons to believe that $i_{58}$ during the first 
migration stage was small, and likely smaller than the current value ($i_{58}\simeq0.05^\circ$). 
If so, ${\rm d}s_8/{\rm d}t>0.05$ arcsec yr$^{-1}$ Myr$^{-1}$ would imply that the crossing of $p_5 = s_8$ 
happened fast enough such that Jupiter's obliquity remained low ($\theta_5<3^\circ$; VN15). 
For comparison, ${\rm d}s_8/{\rm d}t \sim 0.1$ arcsec yr$^{-1}$ Myr$^{-1}$ during the 
first migration stage in NM12. This shows that the NM12 model does not violate the Jupiter obliquity constraint. 

To obtain $\theta_5 \simeq 3^\circ$ from the $p_5 = s_8$ resonance crossing, $i_{58}$ would need to be 
significant. For example, assuming that $i_{58}=0.025^\circ$, i.e. about half of its current value,  
a very slow migration rate with ${\rm d}s_8/{\rm d}t \simeq 0.014$ arcsec yr$^{-1}$ Myr$^{-1}$ would 
be required (VN15), which is well below the expectation from the NM12 model. It thus seems more likely that 
Jupiter's obliquity emerged when $p_5$ approached $s_7$ near the end of planetary migration (Ward \& Canup 2006). For that to work, however, 
the precession constant $\alpha_5$ would have to be significantly larger than $\alpha_5=2.77$ arcsec yr$^{-1}$ 
suggested by Helled et al. (2011). For example, if $\theta_5 < 1^\circ$ before the system approached 
the $p_5 = s_7$ resonance, then $\alpha_5=2.93$-2.95 arcsec yr$^{-1}$ would be needed to obtain 
$\theta_5 \simeq 3^\circ$ (VN15). This constitutes an interesting prediction that will be testable by 
the Juno mission. 
\section{TERRESTRIAL PLANETS}
The principal interaction between the terrestrial and giant planets during planetary migration occurs 
through their {\it secular} coupling. In brief, for a non-resonant system of planets with masses $m_i$ and 
semimajor axes $a_i$, where index $i$ goes from 1 (Mercury) to 8 (Neptune), the secular coupling can be 
described by the Laplace-Lagrange equations. Denoting $(h_i,k_i)=e_i(\sin \varpi_i, \cos \varpi_i)$,
where $e_i$ and $\varpi_i$ are the eccentricity and perihelion longitude of the $i$th planet, the 
Laplace-Lagrange equations admit general solutions with $h_i=\Sigma_j e_{ij} \sin(g_j t + \phi_j)$ and 
$k_i=\Sigma_j e_{ij} \cos(g_j t + \phi_j)$, where $g_j$ are eight eigenfrequencies, and $e_{ij}$ and $\phi_j$ 
are the amplitudes and phases that can be obtained by solving an eigenvalue problem. Similarly, defining 
$(q_i,p_i)=i_i(\sin \Omega_i, \cos \Omega_i)$, where $i_i$ and $\Omega_i$ are the inclination and nodal 
longitude with respect to the invariant plane, it can be shown that $q_i=\Sigma_j i_{ij} \sin(s_j t + \psi_j)$ 
and $p_i=\Sigma_j i_{ij} \cos(s_j t + \psi_j)$, where $s_j$, $i_{ij}$ and $\psi_j$ are eigenfrequencies, 
amplitudes and phases. 

When considering the secular evolution of an isolated system, the semimajor axes of planets are 
constant, and the total angular momentum is conserved. The Angular Momentum Deficit (AMD), defined as
${\rm AMD}=\Sigma_i m_i n_i a_i^2(1-(1-e_i^2)^{1/2} \cos i_i)$, where $n_i$ is the orbital frequency of 
the $i$th planet, is an integral of motion (Laskar 1996). It physically corresponds to the 
angular momentum that needs to be added to make all orbits perfectly circular and coplanar. Using the solution
of the Laplace-Lagrange equations discussed above, AMD can be partitioned into conserved quantities 
$C_j$ and $D_j$ that describe the distribution of the AMD among different eccentricity and inclinations modes 
of each planet (Agnor \& Lin 2012, hereafter AL12). Moreover, the modal amplitudes $C_j$ and $D_j$ are 
constant if $a_j$ change slowly, except if the system evolves near a {\it secular resonance} such that $g_j-g_k=0$ or $s_j-s_k=0$. 

When a secular resonance occurs, the partitioning of the AMD between different modes may change. Because 
there is much more AMD in modes with $j \geq 5$ than in $j\leq4$ (mainly due to the large masses of the 
outer planets), Jupiter and Saturn represent a practically unlimited source of AMD that, if even partially 
transferred to the terrestrial planets, will make their orbits very eccentric and inclined. The orbits 
could then cross each other, leading to collisions between the terrestrial planets.  

These considerations constrain the evolution of the secular modes of the outer planets, mainly $g_5$, from their 
effects on the terrestrial planets. 
For example, using the initial configuration of planets from Hahn \& Malhotra (1999), $P_{\rm 6}/P_{\rm 5}=2.06$, 
where $P_{\rm 5}$ and $P_{\rm 6}$ are the orbital periods of Jupiter and Saturn. The initial orbits are therefore 
just wide of the 2:1 resonance, and $g_5>g_1$ and $g_5>g_2$ (e.g., Fig. 4 in AL12). The present 
value of $g_5$ is 4.24 arcsec yr$^{-1}$, while $g_1=5.86$ arcsec yr$^{-1}$ and $g_2=7.42$ arcsec yr$^{-1}$. 
Therefore, $g_5<g_1$ and $g_5<g_2$ in the present Solar System. This means that the $g_1=g_5$ and $g_2=g_5$ 
resonances must be crossed. The same 
resonances occur in the original Nice model (Tsiganis et al. 2005), but their consequences for 
the terrestrial planets were not understood in 2005.

More recent studies show that the slow crossing of the secular resonances would produce excessive excitation 
and instabilities in the terrestrial planet system (AL12, Brasser et al. 2009). For example, starting from 
the initial $e_{22}=0.01$, and requiring that the final $e_{22}<0.03$ (the present value is $e_{22}=0.018$), 
AL12 did not find {\it any} cases that would satisfy this constraint if the assumed characteristic 
migration timescale $\tau > 0.15$~Myr. In contrast, the timescale of planetesimal-driven migration 
is much longer, with the most quoted values $\tau \geq 5$ Myr (e.g., Hahn \& Malhotra 1999, Gomes et al. 2004, 
Tsiganis et al. 2005, NM12, Nesvorn\'y 2015a). This problem could be resolved if $e_{55} \simeq 0$, because 
the strength of the secular resonances involving the $g_5$ frequency scales with the amplitude $e_{55}$, but 
this would not work either, because it would leave unexplained why $e_{55}=0.044$ now.

\begin{figure}[t]
\includegraphics[width=4.0in]{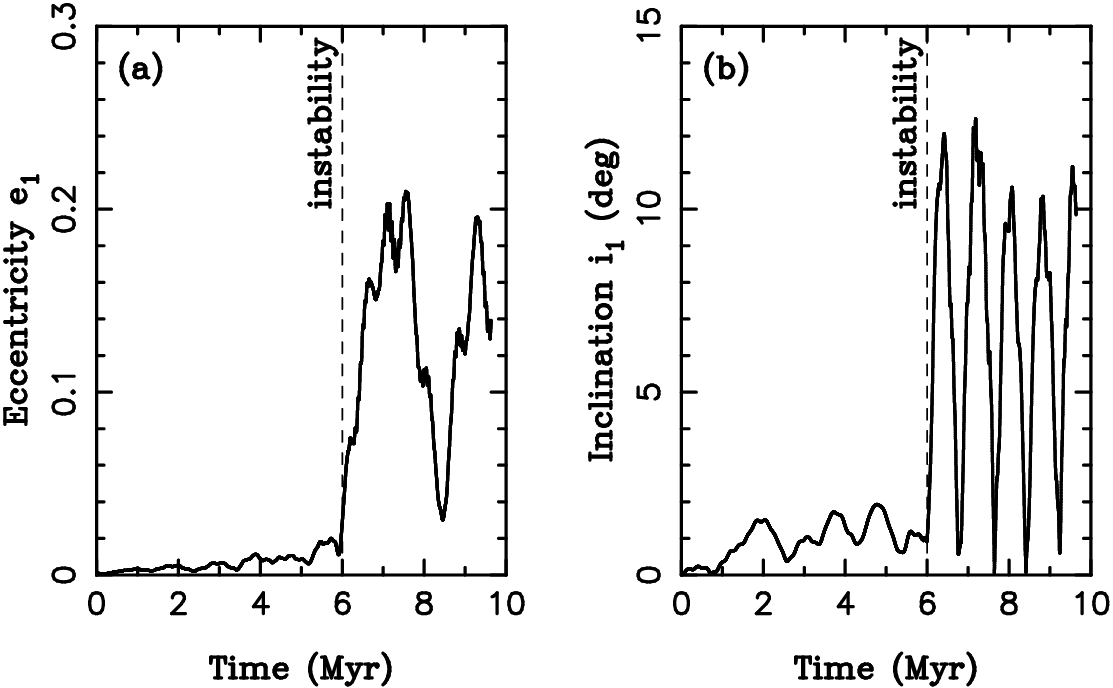}
\caption{Excitation of Mercury's eccentricity (left) and inclination (right) during the dynamical instability
in the outer Solar System. The orbits of outer planets are shown in \textbf{Figure \ref{nice2}}. The instability
happened about 6 Myr after the start of the simulation. The initial AMD of the terrestrial planets was assumed to 
be zero. During the instability, $e_1$ was excited by the $g_1=g_5$ resonance and $i_1$ was excited by 
the $s_1=s_7$ resonance (Roig et al. 2016). The final mean eccentricity and mean inclination were 0.16 and 6.3$^\circ$, 
in a good match to the present orbit of Mercury.}   
\label{terpla}
\end{figure}

As $g_5$ is mainly a function of the orbital separation between Jupiter and Saturn, the constraints from the terrestrial 
planets can be approximately defined in terms of $P_{\rm 6}/P_{\rm 5}$. According to AL12, this 
ratio needs to evolve from $<$2.1 to $>$2.3 in $<$0.15~Myr, which can be achieved, for example, if planetary encounters 
with an ice giant scattered Jupiter inward and Saturn outward (Brasser et al. 2009). This condition has been used to 
measure the success of the instability simulations in NM12 to identify promising cases, which should not violate the 
terrestrial planet constraint.
 In fact, NM12's criterion on $P_{\rm 6}/P_{\rm 5}$ is only a rough expression of the terrestrial planet constraint.  
AL12 showed that even if the migration is very rapid ($\tau = 0.1$~Myr), there is still only a 
$\simeq$40\% chance that the final $e_{22}<0.03$ (assuming constant $e_{55}=0.0443$). A question therefore arises whether the 
NM12 models are truly consistent with the terrestrial planet constraint, or not.

Roig et al. (2016) tested several cases from NM12 (see also Brasser et al. 2013), including the one shown in 
\textbf{Figure \ref{nice2}}. The selected cases were required to pass the NM12 criterion on $P_{\rm 6}/P_{\rm 5}$ (Section 5). 
The terrestrial planets were explicitly included in their $N$-body integrations. The initial AMD 
of the terrestrial planet system was assumed to be much lower than the present one to test whether gravitational 
perturbations from the outer planet system can be responsible for the modestly excited orbits of the 
terrestrial planets. 

The results are interesting. In some cases, the eccentricities and inclinations of the terrestrial 
planets become excited to values similar to the present ones. For example, the excitation by $g_1=g_5$ and $s_1=s_7$ 
resonances can explain why Mercury's orbital eccentricity and inclination are large (mean $e_1=0.17$ and mean $i_1=7^\circ$;
\textbf{Figure \ref{terpla}}).
The Mars's inclination is not excited enough perhaps indicating that Mars acquired its inclined orbit 
(mean $i_4 \simeq 4.5^\circ$) before the planetary migration/instability. In other cases, the simulations failed because 
the planetary orbits were excited too much (Roig et al. 2016).

Kaib \& Chambers (2016) emphasized the low probability that the terrestrial planet system remains unchanged during the 
planetary migration/instability. With the 5-planet model from NM12 and the outer planets starting in the (3:2,3:2,2:1,3:2) 
resonant chain, they found that all terrestrial planet survive in only $\sim$2\% of trials and the AMD generated by the 
planetary migration/instability does not exceed the current one in only $\sim$1\% of trials. This is roughly consistent with 
the previous work, because NM12 found that the $P_6/P_5$ criterion was satisfied in $\sim$5\% of cases, and 
Roig et al. (2016) showed that only some of these cases actually satisfy the terrestrial planet constraint.   

The low probability of matching the terrestrial planet constraint is worrisome. The terrestrial planet region may have 
contained additional planets, either inside the orbit of Venus or outside of Earth's orbit. A planet inside Venus's
orbit would presumably be excited together with Mercury, and its collision with Mercury would probably ensue. The collision 
could reduce the AMD of the terrestrial planets and lead to better results. A hit-and-run collision was 
previously suggested to explain why Mercury looks like a iron core of a larger planet (Asphaug \& Reufer 2014). The 
most straightforward solution to this problem, however, is to assume that the planetary migration/instability happened early, 
within $\sim$50 Myr after the dispersal of the protoplanetary gas disk (Section 13). If so, the terrestrial planet formation 
was not completed and the architecture of the terrestrial planet zone may have been radically different.
\section{ASTEROID BELT}
\begin{figure}[t]
\centering
\begin{minipage}{2.2in}
\centering
\includegraphics[width=2.2in]{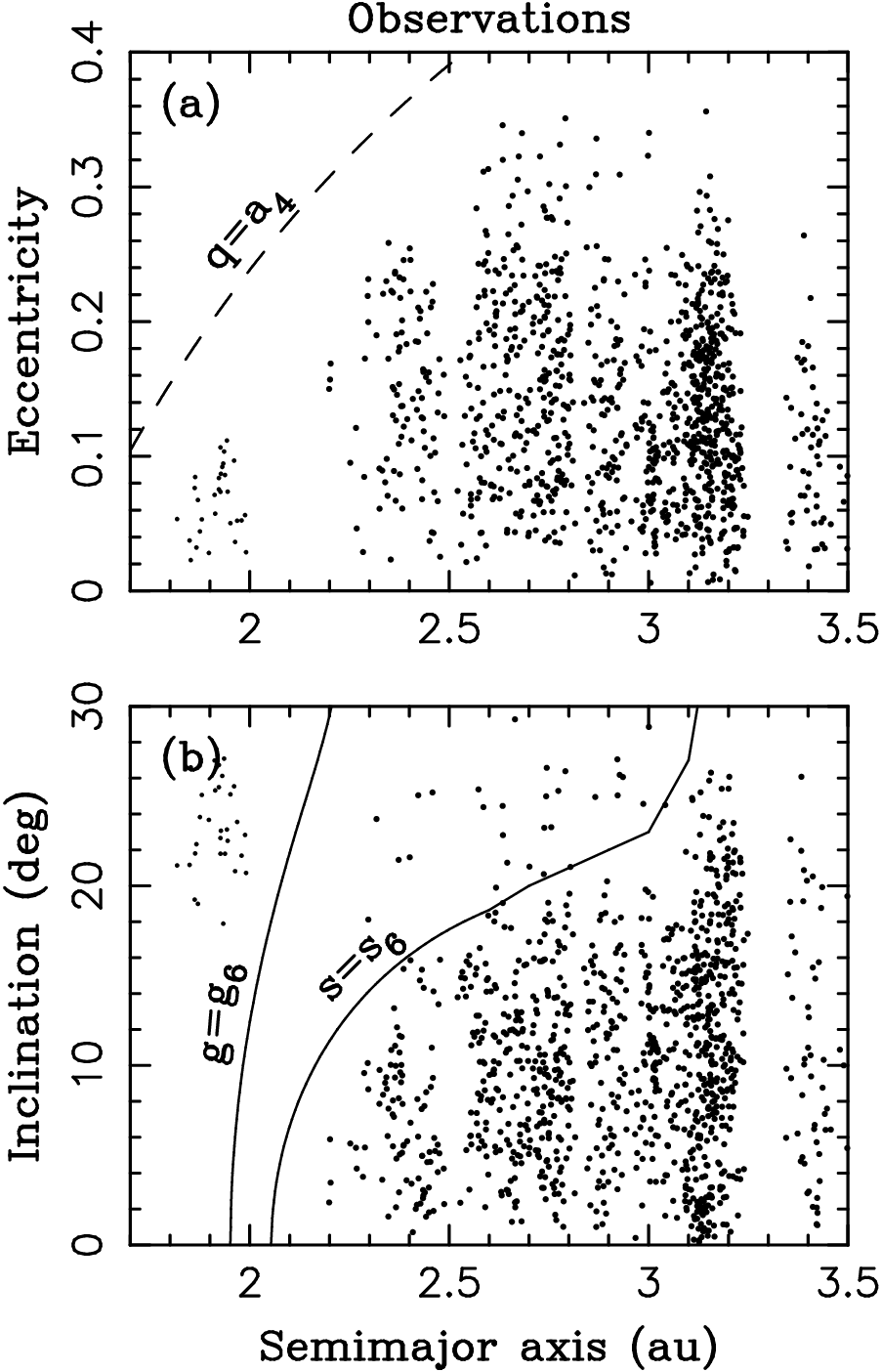}
\end{minipage}
\hspace*{0.1in}
\begin{minipage}{2.2in}
\centering
\includegraphics[width=2.2in]{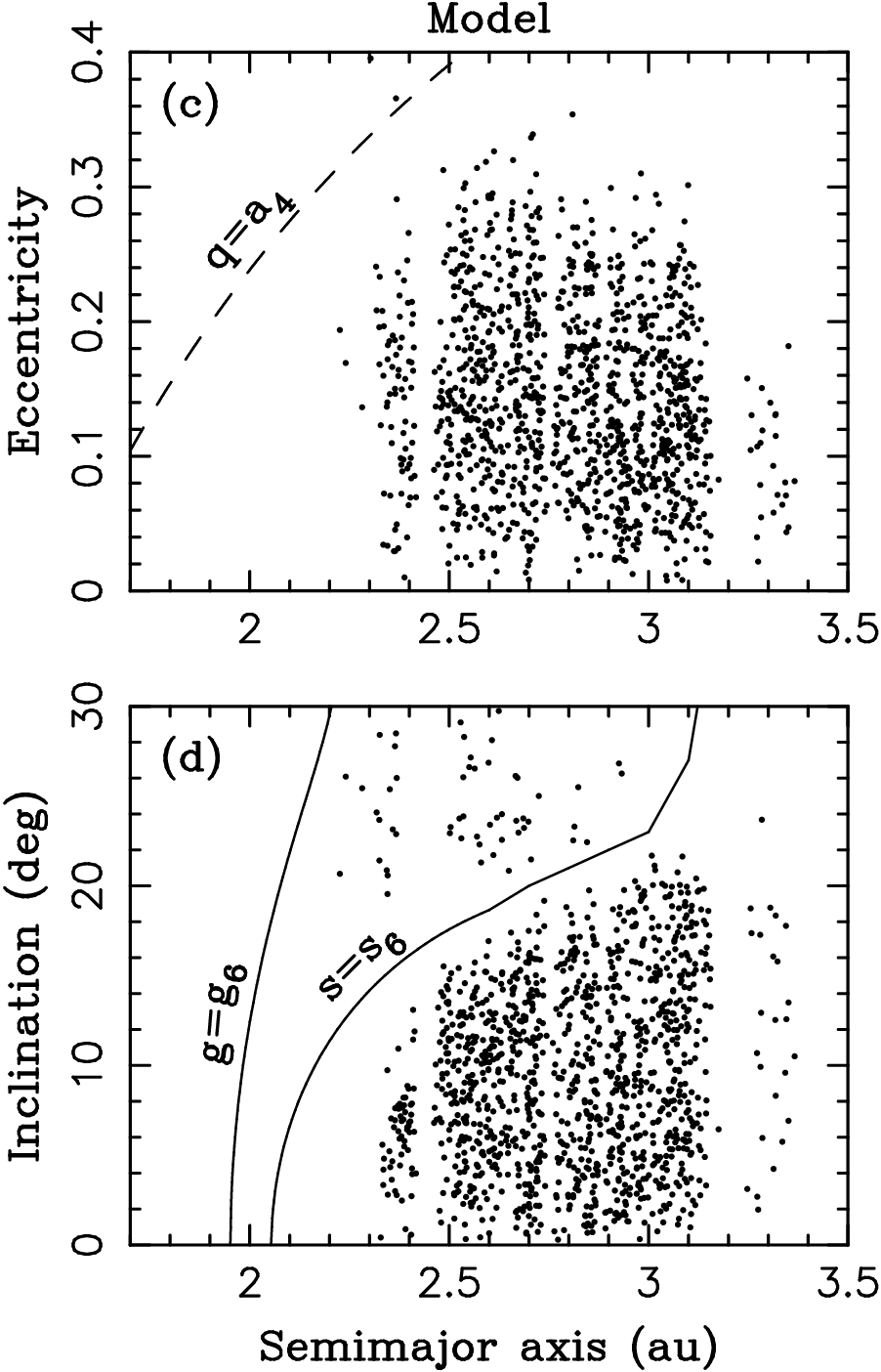}
\end{minipage}
\caption{Left: The orbital distribution of main belt asteroids with diameters $D>30$ km. The 
smaller dots for $a<2$~au highlight Hungarias, which are thought to be a remnant of the E-belt (Bottke 
et al. 2012). Since Hungarias are all smaller than 30 km, here we plot known asteroids with $a<2$ au 
and $D>5$ km. The higher density of orbits at $a\simeq3.1$-3.2 au is a consequence of collisional breakups
of several large asteroids in this region (e.g., Themis and Eos families). 
Right: The distribution of orbits obtained in the jumping-Jupiter model (Nesvorn\'y et al. 2017b). The 
initial inclination distribution was assumed to extend to $i=20^\circ$. The solid lines denote the 
secular resonances $\nu_6$ ($g=g_6$) and $\nu_{16}$ ($s=s_6$). The dashed lines in (a) and (c) is 
where the orbits become Mars crossing.}
\label{aster}
\end{figure}

The orbital distribution of main belt asteroids is carved by resonances. This happens because resonant dynamics
generally increase orbital eccentricities, lead to planet-crossing orbits, and thus tend to remove bodies evolving 
into resonances (the Hilda asteroids in the 3:2 orbital resonance with Jupiter are a notable exception). More specifically, 
the $g=g_6$ and $s=s_6$ resonances (also known as $\nu_6$ and $\nu_{16}$), where $g$ and $s$ are the precession 
frequencies of the proper perihelion and proper nodal longitudes of an asteroid, fall near the current inner boundary 
of the asteroid belt ($a\simeq2$ au for low-inclination orbits; \textbf{Figure~\ref{aster}}). 
The excitation of eccentricities and inclinations in these resonances occurs by processes closely analogous to those 
discussed in Section 7. The orbital resonances, on the other hand, such as 2:1, 3:1, 5:2 and 7:3 with Jupiter (e.g., 
the 2:1 resonance occurs when the orbital period, $P$, is exactly a half of Jupiter's period; $P=11.8/2=5.9$ yr), 
lead to amplified variations of the eccentricity on resonant and secular timescales. The semimajor axis values where 
the resonances occur correspond to gaps in the orbital distribution of asteroids, known as the Kirkwood gaps.

In the early Solar System, when the planetary orbits were different (Sections 2-5), the asteroidal resonances
were at different locations than they are now. For example, with Jupiter on an initial orbit with $a_5\simeq5.8$ au 
(NM12), the 3:1 resonance was at $\simeq$2.8 au, from where it must have moved inwards over the central part of the 
main belt to reach its current location at 2.5 au. Other orbital resonances shifted as well. With Jupiter and Saturn 
in a compact configuration with $P_6/P_5<2.1$, the $\nu_6$ resonance started beyond 4 au, from where it must have moved 
over the whole asteroid belt to $a\simeq2$ au (for $i\sim 0$) when the orbits of Jupiter and Saturn reached their 
current orbital period ratio ($P_6/P_5=2.49$).

To understand this issue, various studies considered the planetesimal-driven migration (Section 2). The studies used 
artificial force terms to induce smooth planetary migration from the initial orbits and placed limits on 
the migration timescale. For example, it has been suggested that an exponential migration $a(t)=a_0+\Delta a [1-\exp(-t/\tau)]$, 
where $a_0$ is the initial semimajor axis, $\Delta a$ is the migration distance ($\Delta a=-0.2$ au for Jupiter 
and $\Delta a=0.8$ au for Saturn), and $\tau \simeq 0.5$ Myr can explain the semimajor axis distribution of main 
belt asteroids (Minton \& Malhotra 2009). In addition, assuming 
$\tau \geq 5$ Myr, which is more consistent with the timescale expected from the planetesimal-driven migration (Section 
11), the sweeping $\nu_{16}$ resonance would excite orbital inclinations to $i>20^\circ$, while inclinations $i>20^\circ$ are 
rare in the present main belt (Morbidelli et al. 2010, Toliou et al. 2016).

The results discussed above therefore imply a very short migration timescale. In this respect, the asteroid belt constraint 
is similar to that obtained from the terrestrial planets (Section 7), except that it applies even if planetary 
migration/instability occurred early.

Very short migration timescales are difficult to obtain from the planetesimal-driven migration, because
that would require a very massive planetesimal disk (Hahn \& Malhotra 1999) and would extract AMD from the outer planets, 
leaving them on more circular orbits than they have now. Instead, it has been suggested that the asteroid constraint can be 
satisfied in the jumping-Jupiter model, where $P_{\rm 6}/P_{\rm 5}$ changes 
in discrete steps with each step corresponding to an encounter of Jupiter or Saturn with an ice giant (Morbidelli et al. 
2010; Section 4). In an idealized version of this model, when Jupiter and Saturn are assumed
to be instantaneously transported from the 3:2 (or 2:1) resonance to their present orbits, the $\nu_6$ and $\nu_{16}$ resonances 
step over the main belt and leave the original orbital distribution of asteroids practically 
unchanged.

The reality is more complicated. Self-consistent simulations of the jumping-Jupiter model show that the radial transport
of planetary orbits is not executed in a single encounter. Instead, both Jupiter and Saturn experience many encounters 
with an ice giant during a period lasting 50,000 to 300,000 years (NM12). The orbital
and secular resonances move in a number of discrete steps over the main belt region and can affect asteroid 
orbits. As for the orbits with $a>2.5$ au, the jumping resonances are found to excite orbital eccentricities; inclinations are 
affected less (Roig \& Nesvorn\'y 2015). 
As for $a<2.5$~au, where both $\nu_6$ and $\nu_{16}$ spend more time, 
the original population is depleted (by a factor of $\sim$10 for $2.1<a<2.5$ au). This can explain why the inner 
part of the belt with $2.1<a<2.5$ au represents only $\sim$1/10 of the total main belt population (\textbf{Figure \ref{aster}}). 
Overall, the main belt loses $\sim$80\% of its original population (Minton \& Malhotra 2010, Nesvorn\'y et al. 2017b). 
The population loss is not large enough to explain the low mass of the main belt ($\simeq$$5\times10^{-4}$ 
$M_\oplus$) when compared to the expectation based on the radial interpolation of the surface density of solids 
between the terrestrial planets and Jupiter's core (Weidenschilling 1977). 
     
In the jumping-Jupiter models investigated so far, the dynamical effects on the orbital eccentricities and 
inclinations are not large enough to explain the general excitation of the asteroid belt (Roig \& Nesvorn\'y 2015).
The processes that excited the belt from the dynamically cold state (that must have prevailed during the 
accretion epoch) therefore most likely {\it predate} the planetary migration/instability (Morbidelli et al. 
2015). For example, the asteroid belt may have become excited (and depleted) before the dispersal of the 
protoplanetary gas disk if Jupiter temporarily moved into the main belt region and scattered asteroids 
around (the GT model; Walsh et al. 2011). In fact, the GT model is known to produce a very broad  
eccentricity distribution (mean $e\sim0.3$-0.4 compared to the present mean $e\sim0.1$). This is not a problem, 
however, because it has been shown that the subsequent dynamical erosion of orbits with $e>0.2$ leads to a narrower 
eccentricity distribution that is more similar to the observed one (Deienno et al. 2016). 
  

Asteroids can be grouped into taxonomic classes based on their reflectance properties. The S-type asteroids show
absorption features similar to the ordinary chondrite meteorites, which are rich in silicates. They are thought
to have formed in the main belt region. The C-type asteroids have featureless neutral spectrum and are likely 
related to carbonaceous chondrites. They are predominant in the central and outer parts of the main belt (2.5-3.3~au)
and are thought to be interlopers from the Jupiter-Saturn zone (Walsh et al. 2011, Kruijer et al. 2017). 
The Cybele asteroids at 3.3-3.7 au, the Hilda asteroids in the 3:2 resonance with Jupiter ($\simeq$3.9 au), and Jupiter 
Trojans are mainly P- (less red spectral slope) and D-types (redder slope). Since Jupiter Trojans are thought to 
have been captured from the outer disk of planetesimals ($a>20$ au; Section 9), the P- and D-type classes are 
probably related to H$_2$0-ice rich comets that formed beyond 20 au. 
Studies show that the outer disk planetesimals can be captured not only as Jupiter Trojans, but also as Hildas, 
Cybeles and in the main belt below 3 au (Levison et al. 2009, Vokrouhlick\'y et al. 2016). 

The fifth planet helps to increase the implantation efficiency into the inner part
of the main belt (Vokrouhlick\'y et al. 2016), where several small P-/D- type asteroids were found (DeMeo et al. 
2015). The mean probability for each outer-disk body to be implanted into the asteroid belt at 2-3.2 au was 
estimated to be $\sim5\times10^{-6}$ (Vokrouhlick\'y et al. 2016). This is consistent with 
the number of large P-/D-type bodies in the belt ($D>150$ km), but represents a significant excess over the 
estimated population of smaller P-/D-types. This problem can be attributed to some physical process that has not 
been included in the existing dynamical studies (thermal or volatile-driven destruction of small P-/D-types 
during their implantation below 3 au, their subsequent collisional destruction, etc.).  
\section{JUPITER TROJANS}
Jupiter Trojans are a population of small bodies with orbits near that of Jupiter. They hug two equilibrium points 
of the three-body problem, known as $L_4$ and $L_5$, with $a \simeq a_5=5.2$~au, $e<0.15$, $i<40^\circ$, and $\delta 
\lambda = \lambda - \lambda_5 \sim \pm 60^\circ$, where $\lambda$ and $\lambda_5$ are the mean longitudes of Trojan
and Jupiter. The angle $\delta \lambda$ librates with a period of $\sim$150 
yr and full libration amplitude, $D$, up to $D\simeq70^\circ$. The color distribution 
of Jupiter Trojans is bimodal with $\sim$80\% of the classified bodies in the red group (red slope similar to that of D 
types in asteroid taxonomy) and $\sim$20\% in the less red group (similar to P types). The distribution of visual albedo 
is uniform with typical values $\simeq$5-7\% (Grav et al. 2011), indicating some of the darkest surfaces in the 
Solar System. 

Morbidelli et al. (2005, hereafter M05) proposed that Jupiter Trojans were trapped in orbits at $L_4$ and $L_5$ by 
{\it chaotic} capture. Chaotic capture takes place when Jupiter and Saturn pass, during their orbital migration, near 
the mutual 2:1 resonance, where the period ratio $P_{\rm 6}/P_{\rm 5} \simeq 2$. The angle $\lambda_{\rm 5}-2\lambda_{\rm 6}
-\varpi$, where $\varpi$ is the mean longitude of either Jupiter or Saturn, then resonates with $\delta \lambda$, 
creating widespread chaos around $L_4$ and $L_5$. Small bodies scattered by planets into the neighborhood of Jupiter's 
orbit can chaotically wander near $L_4$ and $L_5$, where they are permanently trapped once $P_{\rm 6}/P_{\rm 5}$ 
moves away from 2. A natural consequence of chaotic capture is that orbits fill all available space characterized 
by long-term stability, including small libration amplitudes and large inclinations. 

This model resolves a long-standing conflict between previous formation theories that implied $i<10^\circ$ (see Marzari et al. 2002 
and the references therein) and observations that show orbital inclinations up to 40$^\circ$. Attempts to explain large inclinations 
of Trojans by exciting orbits after capture have been unsuccessful, because passing secular resonances and other 
dynamical effects are not strong enough (e.g., Marzari \& Scholl 2000).

\begin{figure}[t!]
\centering
\begin{minipage}{2.4in}
\centering
\includegraphics[width=2.4in]{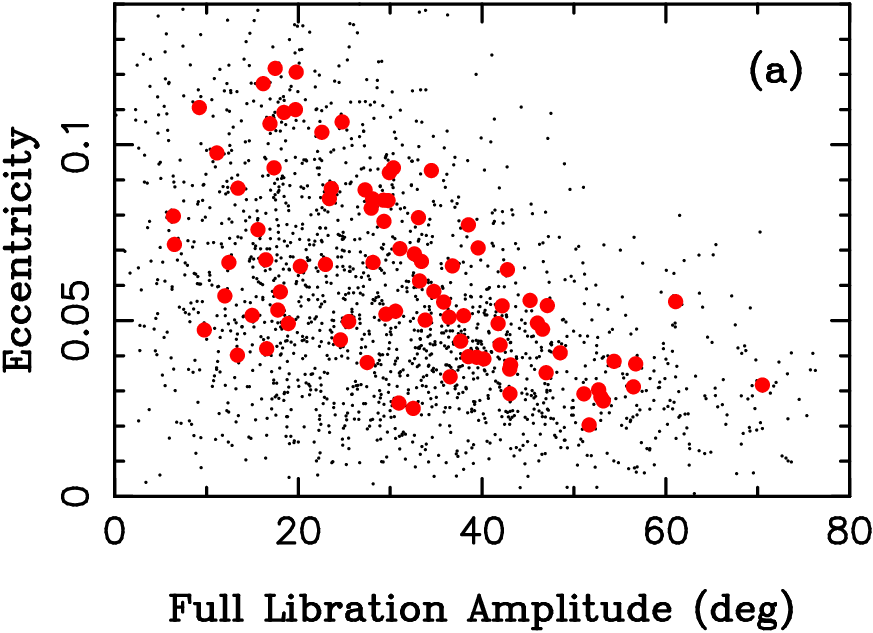}
\end{minipage}
\hspace*{0.1in}
\begin{minipage}{2.4in}
\centering
\includegraphics[width=2.4in]{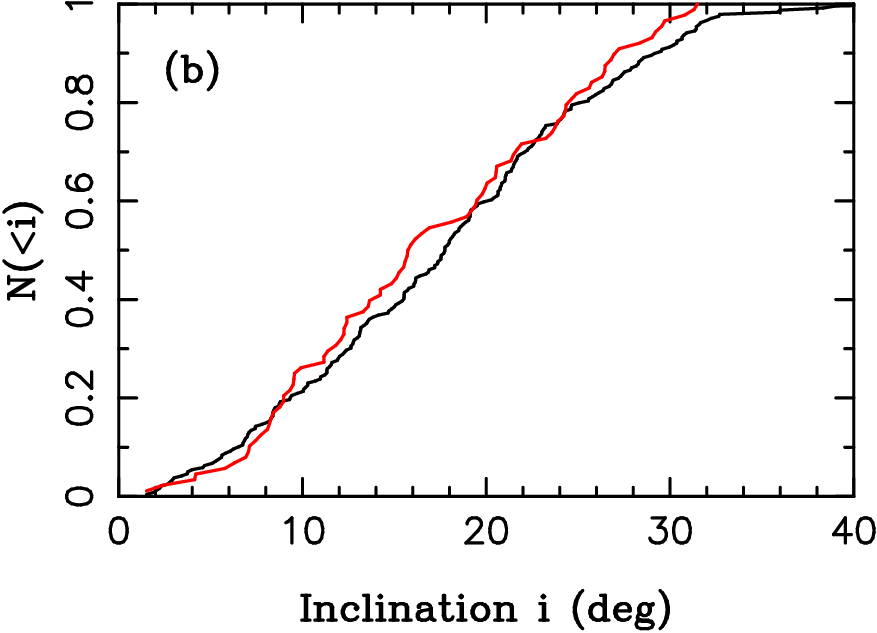}
\end{minipage}
\caption{Left: Orbits of stable Jupiter Trojans obtained in the jumping-Jupiter simulations in NVM13 (red dots). 
The full libration amplitude $D$ corresponds to the angular distance between extremes of $\lambda - \lambda_5$ 
during libration. The black dots show the orbital distribution of known Trojans.
Right: Cumulative inclination distribution of known Jupiter Trojans with absolute magnitude $H<12$ (black line; 
the $H$ cut is done to limit the effect of the observational bias) and bodies captured in the NVM13 model (red 
line).}
\label{trojan}
\end{figure}

M05 placed chaotic capture in the context of the original Nice model (Tsiganis et al. 2005). As we explained 
in Section 4, however, it is now thought that Jupiter and Saturn have {\it not} smoothly migrated over the 2:1 
resonance. Instead, $P_6/P_5$ probably changed from $<$2 to $>$2.3 when Jupiter/Saturn scattered off of an
ice giant. M05's chaotic capture does not work in the jumping-Jupiter model, because the resonances invoked in 
M05~do~not~occur. 

In a follow-up work, Nesvorn\'y, Vokrouhlick\'y \& Morbidelli (2013, hereafter NVM13) tested capture of Jupiter Trojans 
in the jumping-Jupiter model. They found that a great majority of Trojans were captured immediately after 
the closest encounter of Jupiter with an ice giant. As a result of the encounter, $a_5$ changed, sometimes by as 
much as $\sim$0.2 au in a single jump. This radially displaced Jupiter's $L_4$ and $L_5$, released the 
existing Trojans, and led to capture of new bodies that happened to have semimajor axes similar to $a_5$ when 
the jump occurred. NVM13 called this {\it jump} capture. The chaotic capture, arising from the proximity of Jupiter 
and Saturn to the 5:2 resonance, was estimated to contribute only by $\sim$10-20\% to the present population of Jupiter 
Trojans (NVM13).

In principle, both the chaotic and jump capture can produce Trojans from any source reservoir that populated Jupiter's 
region at the time when the orbit of Jupiter changed. For example,  planetesimals from the outer disk 
can be scattered to $\sim$5 au via encounters with the outer planets. The massive outer disk 
($M_{\rm disk}\simeq15$-20 $M_\oplus$) also represents a large source reservoir. Both M05 and NVM13 therefore found that 
an overwhelming majority of Jupiter Trojans were captured from the outer disk. The asteroid contribution 
to Jupiter Trojans is negligible (Roig \& Nesvorn\'y 2015). 

\begin{figure}[t]
\centering
\begin{minipage}{2.4in}
\centering
\includegraphics[width=2.4in]{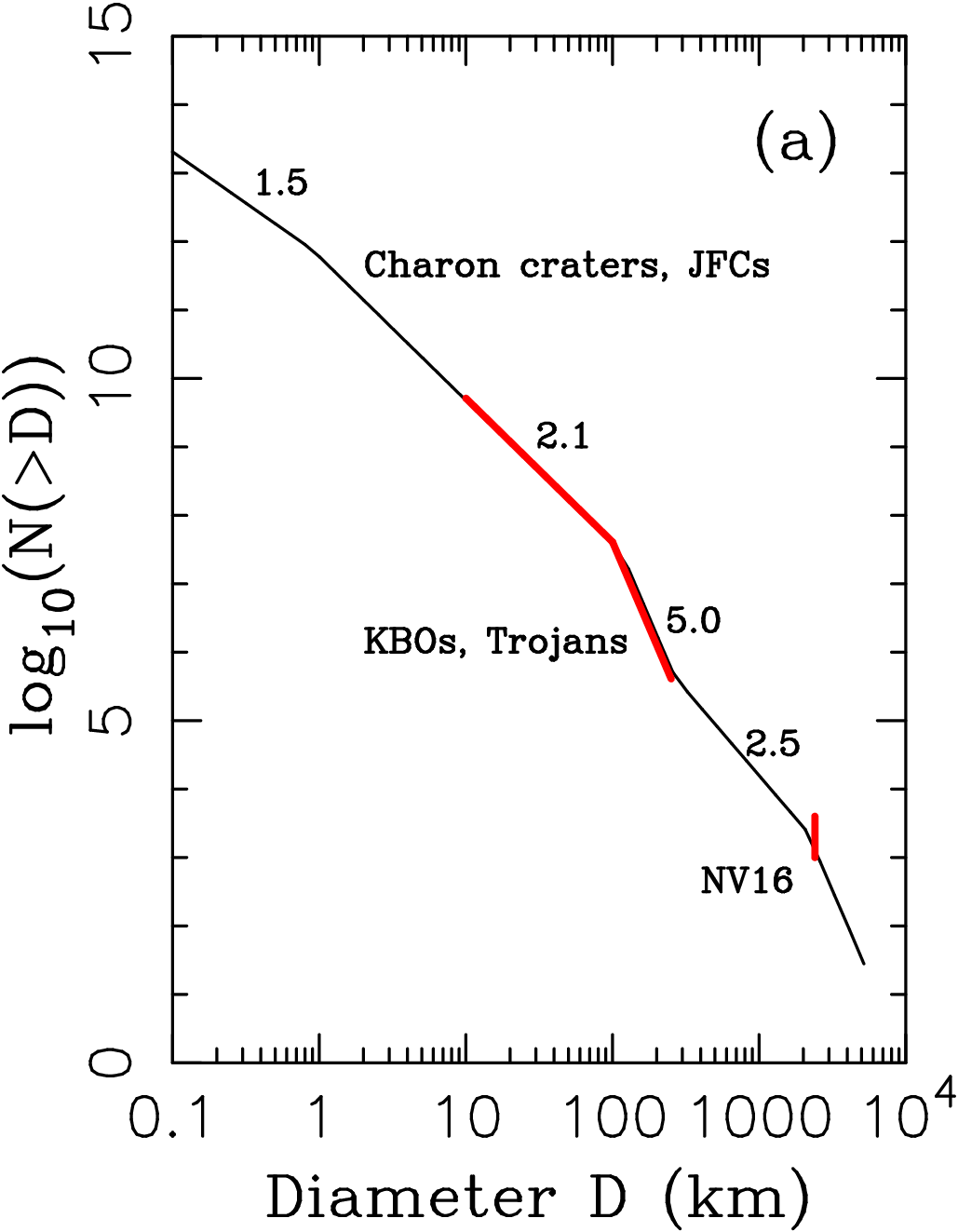}
\end{minipage}
\hspace*{0.1in}
\begin{minipage}{2.31in}
\centering
\vspace*{2.mm} 
\includegraphics[width=2.25in]{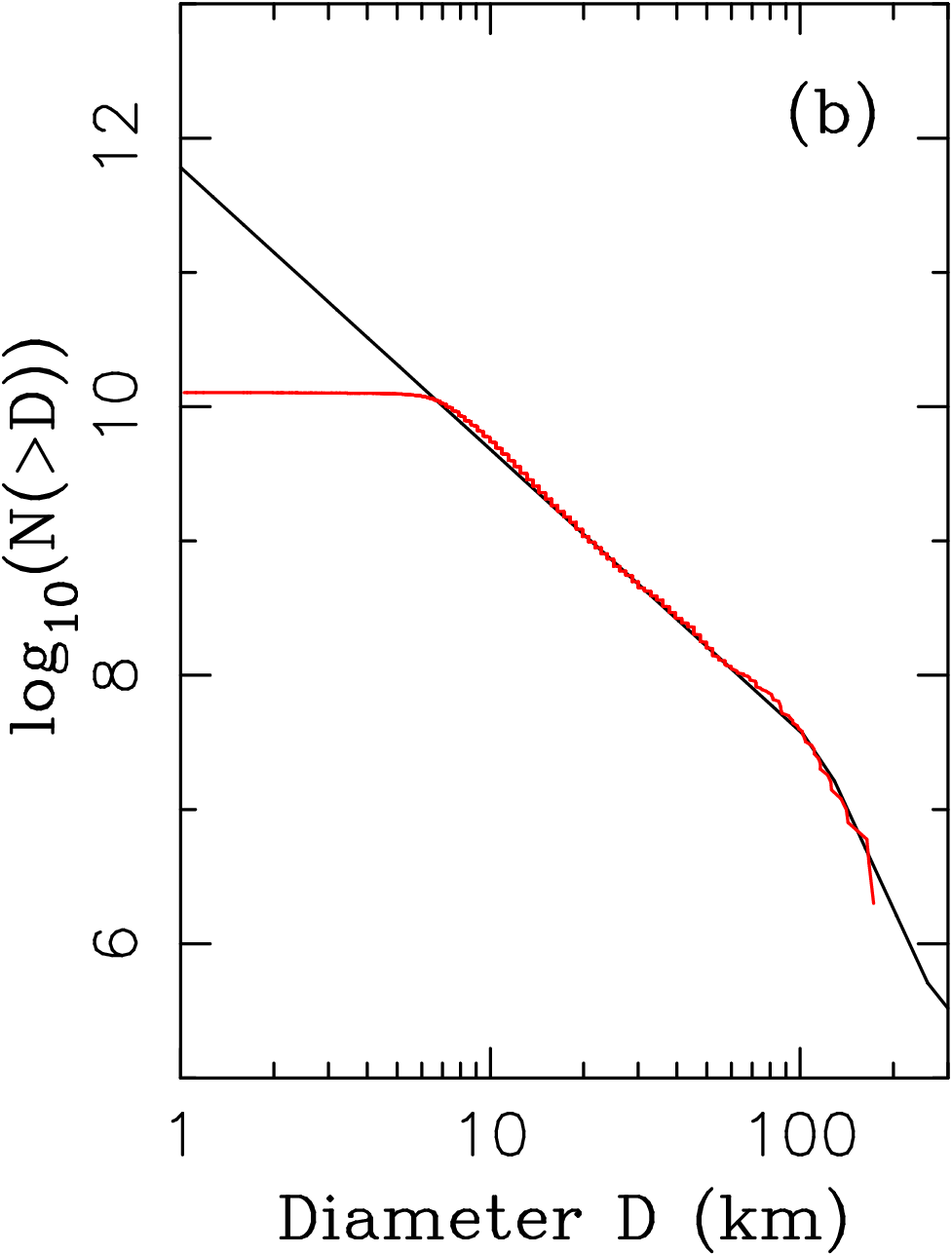}
\end{minipage}
\caption{The calibrated size distribution of the original planetesimal disk below 30 au (panel a). The red color 
denotes various constraints. The distribution for $10<D<300$ km was inferred from observations of Jupiter 
Trojans and KBOs, and from the Jupiter Trojan capture probability determined in NVM13.
Panel (b) zooms in on the distribution of $1<D<250$ km planetesimals. The red line in
panel (b) shows the size distribution of known Jupiter Trojans (the sample is incomplete for $D<10$ km). The 
existence of 1000-4000 Plutos in the original disk inferred in Nesvorn\'y \& Vokrouhlick\'y (2016; labeled as 
NV16 in panel a) requires that the size distribution had a hump at $D>300$ km. The numbers above the reconstructed 
size distribution in panel (a) show the cumulative power index that was used for different segments. The total mass 
of the disk, here $M_{\rm disk}=20$ $M_\oplus$, is dominated by $\simeq$100-km-class bodies.}
\label{sfd}
\end{figure}

The orbital distribution of stable Trojans produced in the NVM13 simulations very closely matches observations 
(\textbf{Figure \ref{trojan}}). The distribution extends down to very small libration amplitudes, small eccentricities 
and small inclinations. These orbits are generally the most difficult to populate in any capture model. The inclination 
distribution of captured objects is wide, reaching beyond 30$^\circ$, just as needed. In the best case, the 
Kolmogorov-Smirnov test (Press et al. 1992) gives 60\%, 68\% and 63\% probabilities that the simulated and 
known distributions of $D$, $e$ and $i$ are statistically the same. The capture probability (as a stable Trojan) was 
found to be $P=(5 \pm 3) \times 10^{-7}$ for each particle in the original planetesimal disk, where the error expresses 
the full range of results obtained in the jumping-Jupiter models tested so far.

Since the capture process is size independent, the size frequency distribution (SFD) of 
Trojans should be a scaled-down version of the transplanetary disk's SFD. 
After capture, the Trojan population underwent collisional evolution as evidenced by the presence of several 
collisional families (e.g, Rozehnal et al. 2016). The collisional evolution has modified the SFD of small bodies but 
left the SFD of large bodies ($D>10$ km) and the total mass of Jupiter Trojans practically unchanged (e.g., 
Wong et al. 2014). Together, these arguments endorse the possibility that the SFD of the present population of 
Jupiter Trojans can be used to reconstruct the SFD of the outer planetesimal disk (Morbidelli et al. 2009b; 
\textbf{Figure \ref{sfd}}).

The magnitude distribution of Jupiter Trojans is known down to $H \sim 17$ (e.g., Yoshida et al. 2017). 
The WISE data show that the SFD has a break at $D_{\rm break} \simeq 100$ km (Grav et al. 2011). The distribution is steep 
for $D>D_{\rm break}$ and shallow for $D<D_{\rm break}$. Below the break, the cumulative SFD can be very well matched by 
a power law $N(>D) = N_{\rm break} (D/D_{\rm break})^\gamma$ with $N_{\rm break}=20$ and $\gamma \simeq 2$. 
The WISE data are incomplete for $D<10$ km, but the measured magnitude distribution indicates that the SFD 
continues with $\gamma \simeq 2.0 \pm 0.2$ below 10 km (Wang \& Brown 2015, Yoshida et al. 2017). This particular 
shape of SFD is thought to have been established by accretional and collisional processes before Trojan capture.  

It has been estimated that the total mass of Jupiter Trojans $M_{\rm JT}\sim10^{-5}$ $M_\oplus$ (M05, Vinogradova \& 
Chernetenko 2015). From WISE data, we have that $M_{\rm JT} \simeq 7.5\times10^{-6}$ M$_\oplus$ (this assumes bulk density 
$\rho=1$ g cm$^{-3}$; Marchis et al. 2014). With $M_{\rm JT}\simeq(0.75$-$1)\times10^{-5}$ M$_\oplus$, it can be  
estimated that the planetesimal disk mass $M_{\rm disk} \sim (0.75$-$1) \times 10^{-5}/(5\times10^{-7})=15$-20 M$_\oplus$. 
This is consistent with $M_{\rm disk} = 15$-20~M$_\oplus$ inferred from the migration/instability simulations (NM12).

The massive outer disk at $\sim$20-30 au was also the source of various KBO populations (Section 11), 
indicating that Jupiter Trojans and KBOs are siblings. 
Indeed, they share the same SFD with a break at $\simeq$100 km (Fraser et al. 2014, Adams et al. 2014). The bulk 
density of Patroclus and Hector, both Jupiter's Trojans, was determined to be $\rho=0.8$-1 g cm$^{-3}$ (Marchis et 
al. 2006, Buie et al. 2015), which is suggestive of high H$_2$O ice content and/or high 
porosity.
The Patroclus and (18974) 1998 WR21 equal-size binaries are probably rare survivors of a much larger population of 
binaries in the outer disk (most binaries were presumably dissociated by collisions and planetary encounters). 
In summary, Jupiter Trojans may represent the most readily accessible repository of Kuiper belt material. 
The NASA Lucy mission, to be launched in 2021, will explore this connection (Levison et al. 2016).   
\section{REGULAR AND IRREGULAR MOONS}
The standard model for the formation of large {\it regular} moons (the Galilean satellites and Titan) is that they 
formed by accretion in circumplanetary disks (Peale 1999). At least some of the mid-sized regular 
moons of Saturn may have formed later during the viscous spreading of young massive rings (Charnoz et al. 
2010). These models cannot be applied to the {\it irregular} moons (see Nicholson et al. 2008 and the references 
therein), because: (i) they are well separated from the regular satellite systems, making it unlikely that 
they formed from the same circumplanetary disk; (ii) their eccentricities, in general, are too large to have 
been the result of simple accretion; and (iii) most of them follow retrograde orbits, so they could not have 
formed in the same disk/ring as the prograde regular satellites. 

The irregular satellites have been assumed to have been captured by planets from heliocentric orbits: 
(1) via dissipation of their orbital energy by gas drag (e.g., \'Cuk \& Burns 2004), (2) by collisions with stray planetesimals, (3) by 
`pull-down' capture, in which the planet's gradual growth leads to capture, or (4) by an exchange reaction when 
a binary enters the planet's Hill sphere, dissolves, and one component ends in a planetocentric orbit. These 
models raise important questions that need to be addressed in more detail. For example, model (4), while 
certainly plausible for capture of Neptune's moon Triton (Agnor \& Hamilton 2006), has a capture efficiency 
about 2-3 orders of magnitude too low to explain the observed population of irregular satellites and produces 
a peculiar orbit distribution of captured objects (Vokrouhlick\'y et al. 2008).

A follow-up work pointed out a serious problem with capture of the irregular satellites by the 
gas-assisted and other mechanisms at early epochs: These early-formed distant satellites are efficiently 
removed at later times when large planetesimals (Beaug\'e et al. 2002) and/or planet-sized bodies sweep through 
the satellite systems during migration of the outer planets in the planetesimal disk. This is especially clear 
in the instability models discussed in Sections 3-5, where planetary encounters occur. Therefore, while 
different generations of irregular satellites may have existed at different times, most irregular satellites 
observed today were probably captured relatively late. 

\begin{figure}[t]
\centering
\begin{minipage}{2.0in}
\centering
\includegraphics[width=2.0in]{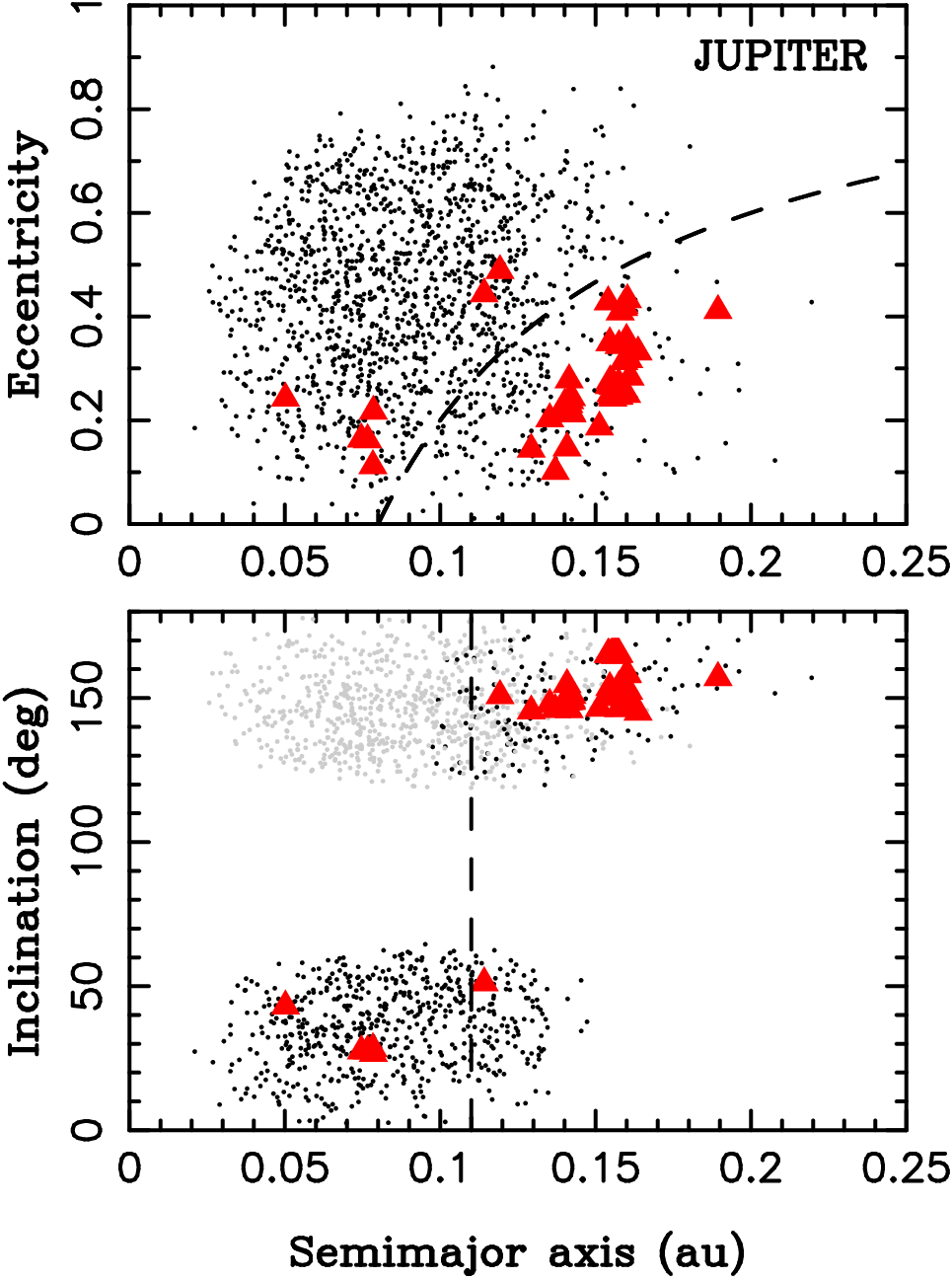}
\end{minipage}
\begin{minipage}{3.0in}
\centering
\includegraphics[width=3.0in]{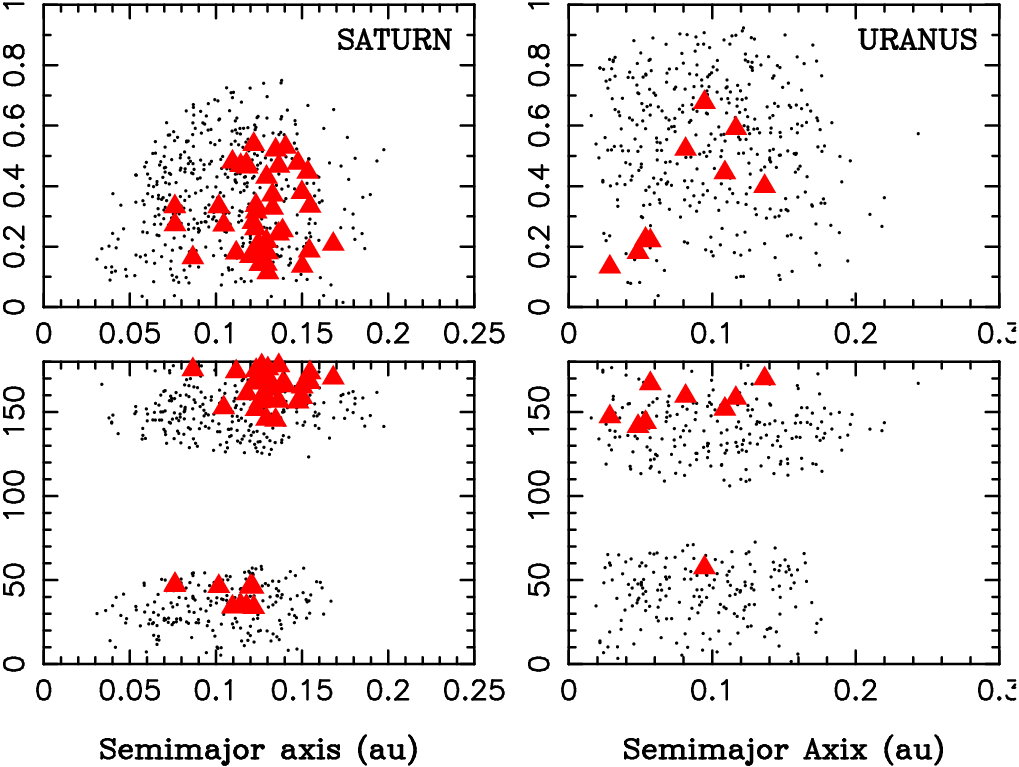}
\end{minipage}
\caption{Left: Orbits of captured satellites (dots) and known irregular satellites at Jupiter 
(red triangles). The dashed line in the top panel denotes $q=a(1-e)=0.08$~au, which is an approximate
limit below which the population of small retrograde satellites becomes strongly depleted by collisions 
with Himalia. The depleted orbits with $i>90^\circ$ and $q<0.08$~au are shown by gray dots in the bottom left 
panel. The dashed line in the bottom panel shows the boundary value $a=0.11$ au below which the collisions 
with Himalia should remove more than 50\% of small retrograde satellites. Right: Orbits of irregular satellites 
at Saturn and Uranus. The results for Neptune, not shown here, can be found in NVM14.}   
\label{isats}
\end{figure}

To circumvent these problems, Nesvorn\'y et al. (2007) suggested that the observed irregular satellites were 
captured from the heliocentric orbits during the time when fully-formed outer planets migrated in the  
planetesimal disk. They considered the original Nice model (Tsiganis et al. 2005). According 
to this model, Saturn and the ice giants repeatedly encounter each other before their orbits get stabilized.
The encounters between planets remove any distant satellites that may have initially formed at Saturn, 
Uranus and Neptune by gas-assisted capture (or via a different mechanism). A new generation of satellites 
is then captured from the background planetesimal disk during planetary encounters. Capture happens when
the trajectory of a background planetesimal is influenced in such a way by the approaching planets that the 
planetesimal ends up on a bound orbit around one of them, where it remains permanently trapped when planets 
move away from each other.

Modeling this mechanism in detail, Nesvorn\'y et al. (2007) found that planetary encounters can create satellites on 
distant orbits at Saturn, Uranus and Neptune with orbital distributions that are broadly similar to those observed. Because 
Jupiter does not generally participate in planetary encounters in the original Nice model, however, the proposed mechanism 
was not expected to produce the irregular satellites at Jupiter. Things changed when the jumping-Jupiter 
model was proposed (Sections 4 and 5), because encounters of Jupiter with an ice giant 
planet is the defining feature of the jumping-Jupiter model. This offered an opportunity to develop a 
unified model where the irregular satellites of {\it all} outer planets are captured by the
same mechanism (with similar capture efficiencies at each planet). This is desirable because the populations 
of irregular satellites at different planets are roughly similar (once it is accounted for the observational 
incompleteness; Jewitt \& Sheppard 2005). 

Nesvorn\'y, Vokrouhlick\'y \& Morbidelli (2014a, hereafter NVM14) studied the capture of irregular satellites
in the five-planet models from NM12 (Section 5). They found that the orbital distribution of bodies captured during 
planetary encounters provides a good match to the observed distribution of the irregular satellites at Jupiter,
Saturn, Uranus and Neptune (\textbf{Figure~\ref{isats}}).  The capture efficiency at Jupiter was found to be 
$(1.3$-$3.6)\times10^{-8}$ for each planetesimal in the original outer disk. The calibration of the outer disk 
from Jupiter Trojans (Section~9) implies that there were $\simeq 6\times 10^9$ $D>10$ km planetesimals 
in the outer disk. Therefore, Jupiter's encounters with the ejected ice giant should produce $\simeq80$-220 
$D>10$ km irregular satellites at Jupiter (NVM14). For comparison, only 10 $D>10$ km irregular satellites are 
known and this sample is thought to be complete. 

The initially large population of captured satellites are expected to be reduced by disruptive 
collisions among satellites (Bottke et al. 2010). The results of the collisional cascade modeling imply a very 
shallow SFD slope for $D \sim 10$ km, exactly as observed. The satellite families provide a direct evidence
for disruptive collisions of satellites (Nesvorn\'y et al. 2003, Sheppard \& Jewitt 2003). Collisions are also 
thought to be responsible for the observed asymmetry between the number of prograde (1 object) and retrograde 
(11 objects) irregular moons at Uranus. The asymmetry arises when the largest moon in the captured population 
eliminates smaller irregular moons that orbit the planet in the opposite sense (Bottke et al. 2010). 

The {\it regular} moons of the outer planets also represent an important constraint on the history of planetary 
encounters. This is because the orbits of the regular moons can be perturbed by gravity of the passing 
planet. In an extreme case, when very deep encounters between planets occur, the orbits of regular 
moons could be excited and destabilized. Deienno et al. (2014) studied the effects of planetary encounters 
on the Galilean satellites in several migration/instability cases from NM12. They found that the strongest 
constraint on the encounters is derives from the small orbital inclinations of the Galilean moons 
($i<1^\circ$). The inclinations of Galilean moons, if exited to $i>1^\circ$, would not evolve to 
$i<1^\circ$ by tidal damping over 4.5 Gyr. Thus, a strong excitation of inclinations during encounters 
must be avoided.   

It has been determined that the largest orbital perturbations occur during a few deepest encounters
(Deienno et al. 2014; the irregular satellites, instead, are captured by many encounters including the distant ones; NVM14). 
The simulation results imply that the encounters with the minimum distance $d<0.02$ au must avoided, and 
the encounters with $0.02<d<0.05$ au cannot be too many (for reference, Callisto has $a\simeq0.012$ au).
Roughly 50\% of NM12 instability cases that satisfy the A-D criteria (Section 5) also satisfy this constraint. 
Interestingly, the distant encounters of Saturn with an ice giant could have excited the Iapetus's inclination
to its current value ($i \simeq 8^\circ$ with respect to the local Laplace plane) while leaving its 
eccentricity low (Nesvorn\'y et al. 2014b). 

The regular satellites of Uranus are a very sensitive probe of planetary encounters. This is because
the most distant of these satellites, Oberon, has only $\simeq$0.068$^\circ$ inclination with respect to 
the Laplace surface. Previous works done in the framework of the original Nice model and the jumping-Jupiter 
model with four planets (Deienno et al. 2011; Nogueira et al. 2013) had difficulties to satisfy this 
constraint, because Uranus experienced encounters with Jupiter and/or Saturn in these instability 
models. In the NM12 model, Uranus does not have encounters with Jupiter and Saturn, 
and instead experiences a small number of encounters with a relatively low-mass ice giant. 
Consequently, Oberon's inclination remains below 0.1$^\circ$ in nearly all cases taken from NM12. 
Neptune's regular satellites are less of a constraint, because Triton's orbit is closely bound to Neptune 
and has been strongly affected by tides (Correia et al. 2009).
\section{KUIPER BELT}
\begin{figure}[t]
\centering
\begin{minipage}{2.3in}
\centering
\includegraphics[width=2.3in]{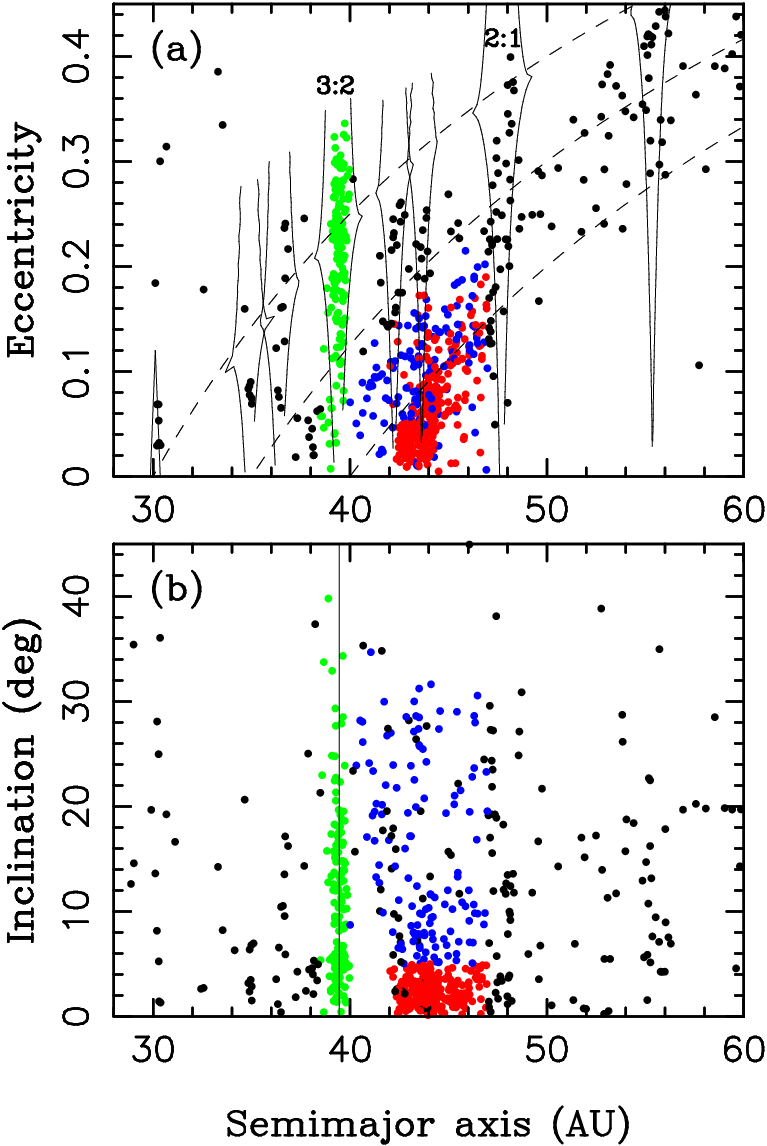}
\end{minipage}
\hspace*{0.1in}
\begin{minipage}{2.3in}
\centering
\includegraphics[width=2.3in]{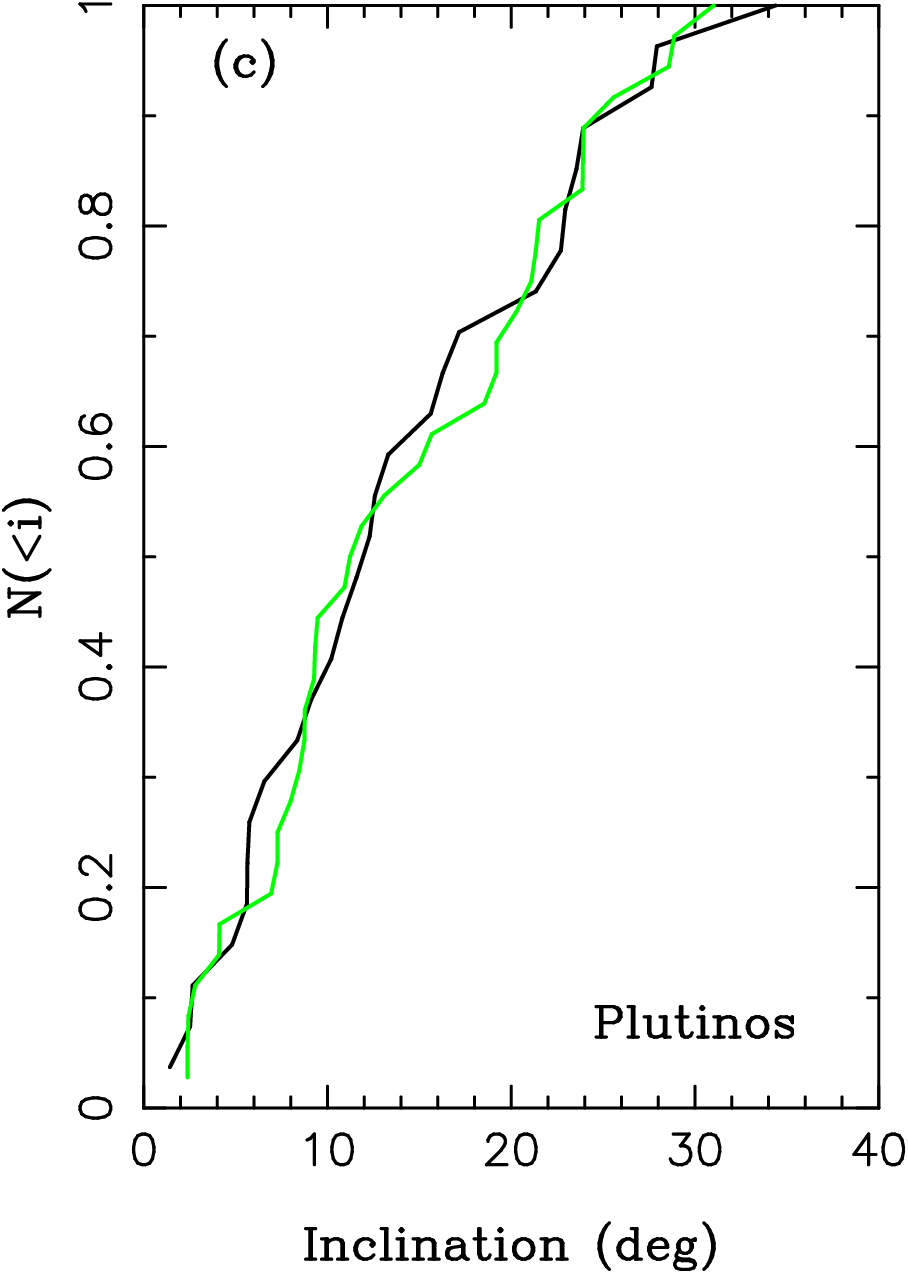}
\end{minipage}
\caption{Left: Orbits of KBOs observed in three or more oppositions. Various dynamical classes 
are highlighted. HCs with $i>5^\circ$ are denoted by blue dots, and CCs with $i<5^\circ$ 
are denoted by red dots. The solid lines in panel (a) follow the borders of important orbital resonances. 
Note the wide inclination distribution of Plutinos (green dots) and HCs in panel (b) with inclinations 
reaching above $30^\circ$. Right: Comparison of the inclination 
distributions obtained in the model ($\tau=30$ Myr, $e_{8,0}=23$ au; Nesvorn\'y \& Vokrouhlick\'y 2016; green line) 
and the Canada-France Ecliptic Plane Survey (CFEPS; Petit et al. 2011) detections (black line). The CFEPS detection 
simulator was applied to the model orbits to have a one-to-one comparison with the actual CFEPS detections.}
\label{kuiper}
\end{figure}

The Kuiper belt is a diverse population of bodies on trans-Neptunian orbits (\textbf{Figure \ref{kuiper}}). 
Based on dynamical considerations, the Kuiper Belt Objects (KBOs) are classified (Gladman et al. 2008) into 
several groups: the resonant populations, classical belt, scattered/scattering disk, and detached objects 
(also known as the fossilized scattered disk). 
The resonant populations are a fascinating feature of the Kuiper belt. They give 
the Kuiper belt an appearance of a bar code with individual bars centered at the resonant orbital periods. 
Pluto and Plutinos in the 3:2 resonance with Neptune (orbital period $\simeq$250~yr) are the largest and 
best-characterized resonant group. The resonant bodies are long-lived, even if $q<Q_{\rm 8}$, where
$Q_{\rm 8}=a_8(1-e_8)\simeq30.4$ au is the aphelion distance of Neptune, because they are phase-protected by 
resonances from close encounters with Neptune. 

The orbits of the scattered/scattering disk objects (SDOs), on the other hand, evolved and keep 
evolving by close encounters with Neptune. These objects tend to have long orbital periods and be
detected near their orbital perihelion when the heliocentric distance is $\sim$30~au. Their neighbors, 
the detached objects, have a slightly larger perihelion distance than the scattered/scattering objects 
and semimajor axes beyond the 2:1 resonance ($a>47.8$~au). The detached KBOs with very large semimajor axes 
($a>150$ au) are sometimes referred to as the extreme SDOs. The observed orbital alignment of extreme 
SDOs has driven the recent interest in the Planet 9 hypothesis (Trujillo \& Sheppard 2014, Batygin 
\& Brown 2016).   

The classical Kuiper Belt, hereafter CKB, is a population of trans-Neptunian bodies dynamically defined as 
having non-resonant orbits with perihelion distances that are large enough to avoid close encounters with Neptune.  
Most known KBOs reside in the main CKB located between the 3:2 and 2:1 resonances with Neptune ($39.4<a<47.8$ au). 
It is furthermore useful to divide the CKB into dynamically ``cold'' and ``hot'' components, mainly because the 
inclination distribution in the CKB is bimodal (Brown 2001, Gulbis et al. 2010), hinting at different dynamical 
origins of these groups. The Cold Classicals (CCs) are often defined as having $i<5^\circ$ and 
Hot Classicals (HCs) as $i>5^\circ$ (\textbf{Figure \ref{kuiper}}). Note that this definition is somewhat arbitrary, 
because the continuous inclination distribution near $i=5^\circ$ indicates that significant mixing between the 
two components must have occurred (e.g., Volk \& Malhotra 2011).

While HCs share many similarities with other dynamical classes of KBOs (e.g., scattered disk, Plutinos), 
CCs have several unique properties. Specifically, (1) CCs have distinctly red colors (e.g., Tegler \& 
Romanishin 2000) that may have resulted from space weathering of surface ices, such as ammonia (e.g., 
Brown et al. 2011), that are stable beyond $\sim$35 au. (2) A large fraction of 100-km-class CCs are wide 
binaries with nearly equal size components (Noll et al. 2008). (3) The albedos of CCs are generally higher 
than those of HCs (Brucker et al. 2009). And finally, (4) the size distribution of CCs is markedly different 
from those of the hot and scattered populations, in that it shows a very steep slope at large sizes (e.g., 
Bernstein et al. 2004), and lacks very large objects. The most straightforward interpretation of these 
properties is that CCs formed and/or dynamically evolved by different processes than other trans-Neptunian populations.

The complex orbital structure of the trans-Neptunian region with heavily populated resonances, and high 
eccentricities and high inclinations of orbits (\textbf{Figure \ref{kuiper}}), does not represent the dynamical conditions 
in which KBOs accreted. Instead, it is thought that much of this structure appeared as a result of Neptune's migration. 
Following the pioneering work of Malhotra (1993, 1995), studies of Kuiper belt dynamics first considered 
the effects of outward migration of Neptune that can explain the prominent populations of KBOs in the
major orbital resonances (Levison \& Morbidelli 2003, Gomes 2003, Hahn \& Malhotra 2005). With the advent of the 
notion that the early Solar System may have suffered a dynamical instability (Sections 3-5), the focus broadened, with recent 
theories invoking an eccentric and inclined orbit of Neptune (Levison et al. 2008, Batygin et al. 2011, 
Wolff~et~al.~2012, Dawson \& Murray-Clay 2012, Morbidelli et al. 2014). 

The emerging consensus is that HCs formed at $<$30 au, and were dynamically scattered to their current orbits 
by migrating/eccentric Neptune, while CCs formed at $>$40~au and survived Neptune's early `wild days' relatively 
unharmed. The main support for this model comes from the unique properties of CCs, which would be difficult 
to explain if HCs and CCs had similar formation locations (and dynamical histories). For example, the wide binaries 
observed among CCs would not survive scattering encounters with Neptune (Parker \& Kavelaars 2010). 

An outstanding problem with the previous models of Kuiper belt formation (e.g., Hahn \& Malhotra 2005, Levison
et al. 2008) is that the predicted distribution of orbital inclinations of Plutinos and HCs was found to be narrower 
than the one inferred from observations. The inclinations may have been excited before Neptune's migration, but
no such an early excitation process was identified so far. This problem appears to be more likely related to the 
timescale of Neptune's migration. Nesvorn\'y (2015a) performed numerical simulations of Kuiper belt formation starting 
from an initial state with a dynamically cold massive outer disk extending from beyond $a_{\rm N,0}$ to 30 au. 
According to arguments discussed in Sections 5, and the calibration from Jupiter Trojans (Section 9)
the original disk mass was assumed to be $M_{\rm disk}\simeq15$-20 $M_\oplus$.  

In different simulations, Neptune was started with $20<a_{8,0}<30$ au and migrated into the disk on an e-folding 
timescale $1 < \tau < 100$ Myr to test the dependence of the results on the migration range and timescale. A small fraction 
of the disk planetesimals became implanted into the Kuiper belt in the simulations. 
To satisfy the inclination constraint (\textbf{Figure \ref{kuiper}c}), it was found that Neptune's migration must have been slow 
($\tau \geq 10$ Myr) and long range ($a_{8,0} < 25$ au). The models with $\tau < 10$~Myr do not satisfy the inclination 
constraint, because there is not enough time for dynamical processes to raise inclinations. The slow migration of 
Neptune is consistent with other Kuiper belt constraints, and represents an important clue about the original mass of
the outer disk. For example, in the NM12 planetary migration/instability model where the outer disk extends from $\simeq$23 
to 30 au, $\tau \geq 10$ Myr implies $M_{\rm disk} \simeq 15$-20 $M_\oplus$. 

Neptune's eccentricity and inclination are never large in the NM12 models ($e_{\rm N}<0.15$, $i_{\rm N}<2^\circ$), as required 
to avoid excessive orbital excitation in the $>$40 au region, where the CCs formed. 
Simulations show that the CC population was dynamically depleted by only a factor of $\sim$2 (Nesvorn\'y 2015b). This implies 
that the surface density of solids at 42-47 au was $\sim$4 orders of magnitude lower than the surface density needed
to form sizable objects in the standard coagulation model (Kenyon et al. 2008). It is possible that the original surface
density was higher and bodies were removed by fragmentation during collisions (Pan \& Sari 2005), but the presence of 
loosely bound binaries places a strong constraint on how much mass can be removed by collisions (Nesvorn\'y et al. 2011).
Instead, these results suggest that CCs accreted in a low-mass environment (Youdin \& Goodman 2005).
 
A particularly puzzling feature of the CC population is the so-called kernel, a concentration of orbits with $a=44$ au, 
$e\simeq0.05$ and $i<5^\circ$ (Petit et al. 2011). This feature can either be interpreted as a sharp edge beyond which the 
number density of CCs drops, or as a genuine concentration of bodies. If it is the latter, the kernel can be explained if 
Neptune's migration was interrupted by a discontinuous change of Neptune's semimajor axis when Neptune reached $\simeq$27.7~au 
(the jumping-Neptune model; Petit et al. 2011, Nesvorn\'y 2015b). Before the discontinuity happened, planetesimals located 
at $\sim$40~au were swept into the Neptune's 2:1 resonance, and were carried with the migrating resonance outward (Levison 
\& Morbidelli 2003). The 2:1 resonance was at $\simeq$44~au when Neptune reached $\simeq$27.7~au. 
If Neptune's semimajor axis changed by a fraction of an au at this point, perhaps because it was scattered off of 
another planet (NM12), the 2:1 population would have been released at $\simeq$44~au, and would remain there to this day. 
The orbital distribution of bodies produced in this model provides a good match to the orbital properties of the kernel
(Nesvorn\'y 2015b). 

Models with smooth migration of Neptune invariably predict excessively large resonant populations (e.g., Hahn \& Malhotra
2005, Nesvorn\'y 2015a), while observations show that the non-resonant orbits are in fact common (e.g., the classical belt 
population is $\simeq$2-4 times larger than Plutinos in the 3:2 resonance; Gladman et al. 2012). This problem can
be resolved if Neptune's 
migration was {\it grainy}, as expected from scattering encounters of Neptune with massive planetesimals. 
The grainy migration acts to destabilize resonant bodies with large libration amplitudes, a fraction of which ends up on 
stable non-resonant orbits. Thus, the non-resonant--to--resonant ratio obtained with the grainy migration is higher, 
up to $\sim$10 times higher for the range of parameters investigated in Nesvorn\'y \& Vokrouhlick\'y (2016), than in a model 
with smooth migration. The best fit to observations was obtained when it was assumed that the outer planetesimal disk 
below 30 au contained 1000-4000 Plutos. The combined mass of Pluto-class objects in the original disk was thus 
$\sim$2-8 $M_\oplus$, which represents 10-50\% of the estimated disk mass.

Together, the results discussed above imply that Neptune's migration was slow, long-range and grainy, and that Neptune 
radially jumped by a fraction of an au when it reached 27.7 au. This is consistent with Neptune's orbital evolution obtained 
in the NM12 models. Additional constraints on Neptune migration can be obtained from SDOs.
Models imply that bodies scattered outward by Neptune to semimajor 
axes $a>50$ au often evolve into resonances which subsequently act to raise the perihelion distances of detached 
orbits to $q>40$~au (Gomes 2011). The implication of the model with slow migration of Neptune is that 
the orbits with $50<a<100$ au and $q>40$~au should cluster near (but not in) the resonances with Neptune (3:1 at $a=62.6$ au, 
4:1 at $a=75.9$ au, 5:1 at $a=88.0$ au; Kaib \& Sheppard 2016, Nesvorn\'y et al. 2016). The recent detection of several 
distant KBOs near resonances is consistent with this prediction, but it is not yet clear whether most orbits are 
really non-resonant.
\section{COMETARY RESERVOIRS}
Comets are icy objects that reach the inner Solar System after leaving distant reservoirs beyond Neptune and 
dynamically evolving onto elongated orbits with very low perihelion distances (Dones et al. 2015). 
Their activity, manifesting itself by the presence of a dust/gas coma and characteristic tail, is driven by 
solar heating and sublimation of water ice. Comets are short-lived, implying that they must be resupplied from 
external reservoirs (Fern\'andez 1980, Duncan et al. 1988). 

\begin{figure}[t]
\centering
\begin{minipage}{2.1in}
\centering
\includegraphics[width=2.1in]{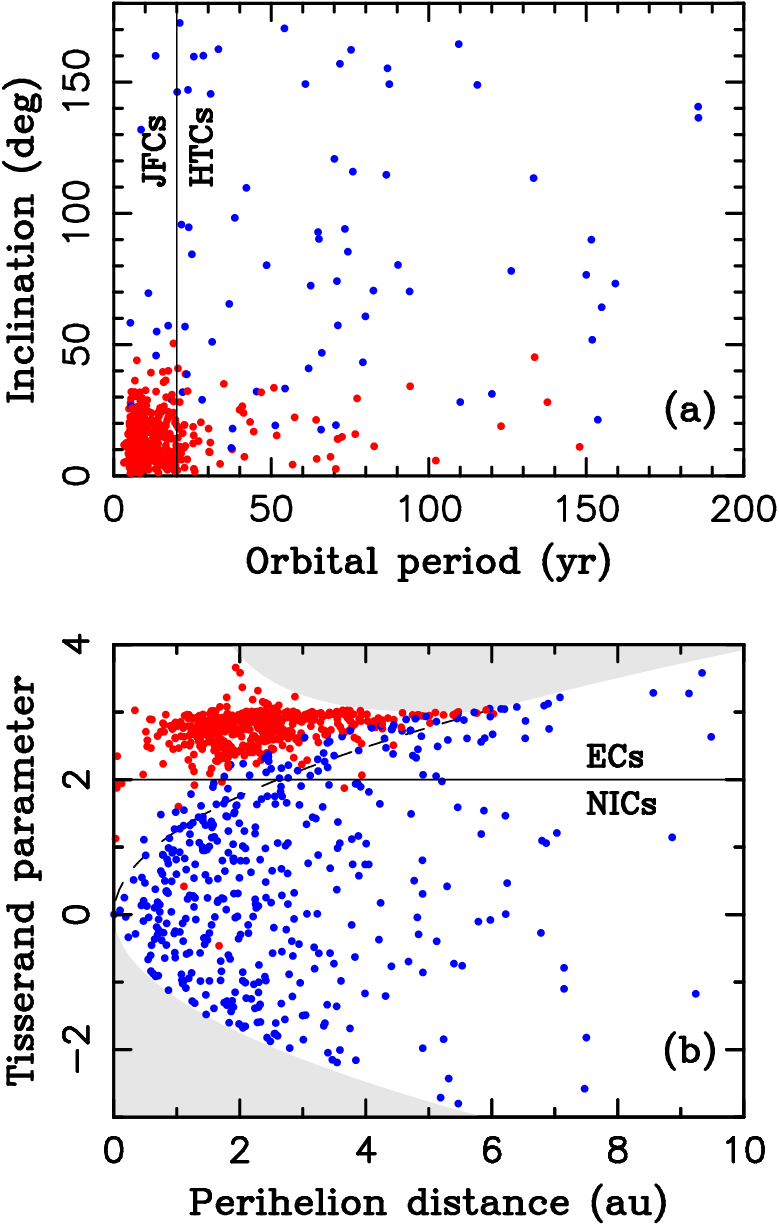}
\end{minipage}
\hspace*{0.0in}
\begin{minipage}{2.85in}
\centering
\includegraphics[width=2.85in]{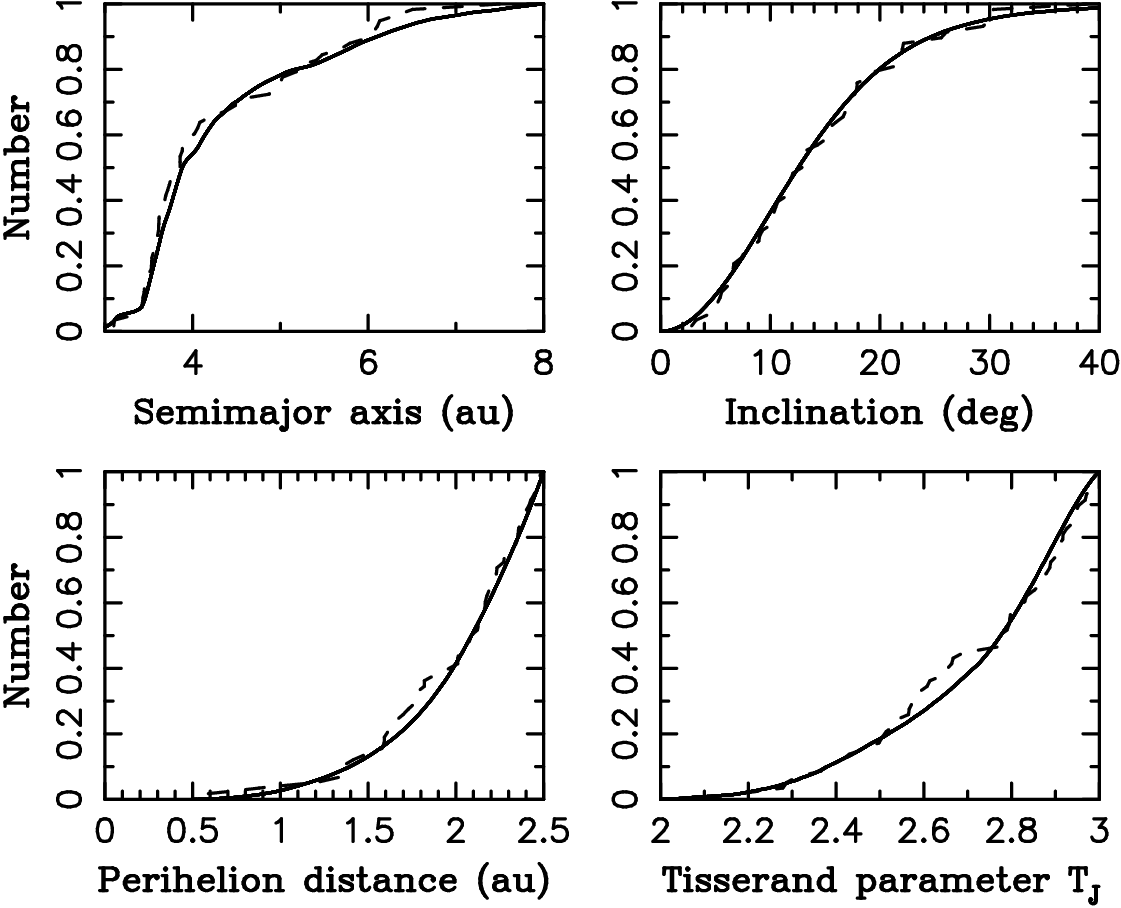}
\end{minipage}
\caption{Left: The orbital distribution of known SPCs. The thin lines show the division between 
JFCs and HTCs (panel a; $P=20$ yr), and between ECs and NICs (panel b; Tisserand parameter 
$T_{\rm J}=2$). The color indicates the relationship between different categories. In panel a, the red dots denote
ECs with $T_{\rm J}>2$, and the blue dots denote NICs with $T_{\rm J}<2$. In panel b, the red 
dots denote JFCs with $P<20$ yr, and the blue dots denote comets with $P>20$ yr and $a<10,$000~au. 
The gray areas in panel b cannot be reached by orbits. The dashed line in panel b is 
$T_{\rm J}=2\sqrt{2q}$, which is an approximate boundary of prograde orbits evolving from 
$a \gg a_{\rm J}$ and $e \sim 1$. Right: The cumulative orbital distributions of ECs with $P<20$ yr, 
$2<T_{\rm J}<3$ and $q<2.5$ au. The model results (solid lines; N17) are compared 
to the distribution of known JFCs (dashed lines). In the model, it was assumed that ECs remain 
active and visible for $N_{\rm p}(2.5)=500$ perihelion passages with $q<2.5$ au.}
\label{comet}
\end{figure}

Levison \& Duncan (1997, hereafter LD97) considered the origin and evolution of ecliptic comets (ECs; see 
\textbf{Figure \ref{comet}} for their relationship to the Jupiter-family comets, JFCs). The Kuiper belt at 30-50 au was assumed 
in LD97 to be the main source reservoir of ECs. Small KBOs evolving onto a Neptune-crossing orbit can be slingshot, 
by encounters with different planets, to very low perihelion distances ($q<2.5$~au), at which point they are 
expected to become active and visible. The new ECs, reaching $q<2.5$~au for the first time, have a narrow inclination 
distribution in the LD97 model, because their orbits were assumed to start with low inclinations ($i<5^\circ$) 
in the Kuiper belt, and the inclinations stay low during the orbital transfer. 

The escape of bodies from the classical KB at 30-50 au is driven by 
slow chaotic processes in various orbital resonances with Neptune. Because these processes affect only 
part of the belt, with most orbits in the belt being stable, questions arise about the 
overall efficiency of comet delivery from the classical KB. Duncan \& Levison (1997), concurrently 
with the discovery of the first SDO; (15874) 1996 TL66,  Luu et al. 1997),
suggested that the scattered disk should be a more prolific source of ECs than the classical KB. 
This is because SDOs can always approach Neptune during their perihelion passages and be scattered by 
Neptune to orbits with shorter orbital periods. 

The Halley-type comets (HTCs) have longer orbital periods and larger inclinations than do most ECs. 
It has been suggested that HTCs evolve into 
the inner Solar System from an inner, presumably flattened part of the Oort cloud (Levison et al. 2001).
This theory was motivated by the inclination distribution of HTCs, which was thought to be flattened 
with a median of $\simeq$45$^\circ$. Later on, the scattered disk was considered as the main source of 
HTCs (Levison et al. 2006). Back in 2006, the median orbital inclination of HTCs was thought to be 
$\simeq$55$^\circ$, somewhat larger than in 2001, but still clearly anisotropic. This turns out to be part of
a historical trend with the presently available data indicating a nearly isotropic inclination distribution 
of HTC orbits (Wang \& Brasser 2014).

The ECs/JFCs and HTCs are also known as the Short-Period Comets (SPCs), defined as bodies showing cometary 
activity and having short orbital periods ($P<200$ yr). The period range is arbitrary, because there is 
nothing special about the boundary at the 200-yr period, and the orbital period distribution of known 
comets appears to continue smoothly across this boundary. With $P<200$ yr, SPCs are guaranteed to have at 
least one perihelion passage in modern history, with many being observed multiple times. This contrasts 
with the situation for the Long-Period Comets (LPCs; $P>200$ yr), which can be detected only if their 
perihelion passage coincides with the present epoch. Disregarding the period cutoff, HTCs and LPCs 
have the common property of having the Tisserand parameter with respect to Jupiter $T_{\rm J}<2$,
and are referred to as the Nearly Isotropic Comets (NICs; LD97 and \textbf{Figure \ref{comet}}). The main reservoir 
of LPCs is thought to be the Oort cloud, a roughly spherical structure of bodies at orbital distances 
$\simeq$$10^4$-$10^5$ au from the Sun.

Our understanding of the origin and evolution of comets is incomplete in part because the presumed source 
populations of trans-Neptunian objects with cometary sizes ($\sim$1-10 km) are not well characterized 
from observations. It is therefore difficult to establish whether there are enough small objects  in any 
trans-Neptunian reservoir to provide the source of comets (e.g., Volk \& Malhotra 2008). To circumvent 
this problem, several recently developed models performed end-to-end simulations in which cometary 
reservoirs are produced in the early Solar System and evolved over 4.5~Gyr (Brasser \& Morbidelli 2013; 
Nesvorn\'y et al. 2017a, hereafter N17). The number of comets produced in the model at the present time 
can then be inferred from the number of comets in the original planetesimal disk, which in turn can be 
calibrated from the number of Jupiter Trojans (Section 9; \textbf{Figure \ref{sfd}}). 
 
This approach, to be reliable, requires that we have a good model for the early evolution of the 
Solar System, which was adopted from NM12 (Section 5). The steady state model of ECs/JFCs obtained in the 
NM12 model can be compared to observations. To do this comparison correctly, as pointed out in LD97, it must be 
accounted for the physical lifetime of active comets (i.e., how long comets remain active). Several 
different parametrizations of the physical lifetime were considered in N17. In the simplest parametrization, 
they counted the number of perihelion passages with $q<2.5$ au, $N_{\rm p}(2.5)$, and assumed that a comet 
becomes inactive if $N_{\rm p}(2.5)$ exceeds some threshold. The threshold was determined by the orbital 
fits to observations. 

The orbital distribution of ECs was well reproduced in the model (\textbf{Figure \ref{comet}}). The nominal 
fit to the observed inclination distribution of JFCs requires, on average, that km-sized JFCs survive 
$N_{\rm p}(2.5)\simeq500$ perihelion passages with $q<2.5$ au. This is consistent with the measured mass loss 
of 67P/Churyumov-Gerasimenko (Paetzold et al. 2016).  To explain the number of known large ECs ($D>10$ km), 
large comets are required to have longer physical lifetimes than small comets. 
The dependence of $N_{\rm p}(2.5)$ on comet size for $D<1$ km is poorly constrained, but the physical
lifetime should drop more steeply than a simple extrapolation from $D>1$~km to $D<1$ km would suggest.
This is because $N_{\rm p}(2.5)<10$ is required to match the ratio of returning-to-new LPCs, which presumably 
have $D<1$ km (Brasser \& Morbidelli 2013). The hypothesized transition to very short physical lifetimes for 
comets below 1 km may be related to the rotational spin-up of small cometary nuclei and their subsequent 
disruption by the centrifugal force.

\begin{figure}[t]
\centering
\begin{minipage}{2.3in}
\centering
\includegraphics[width=2.3in]{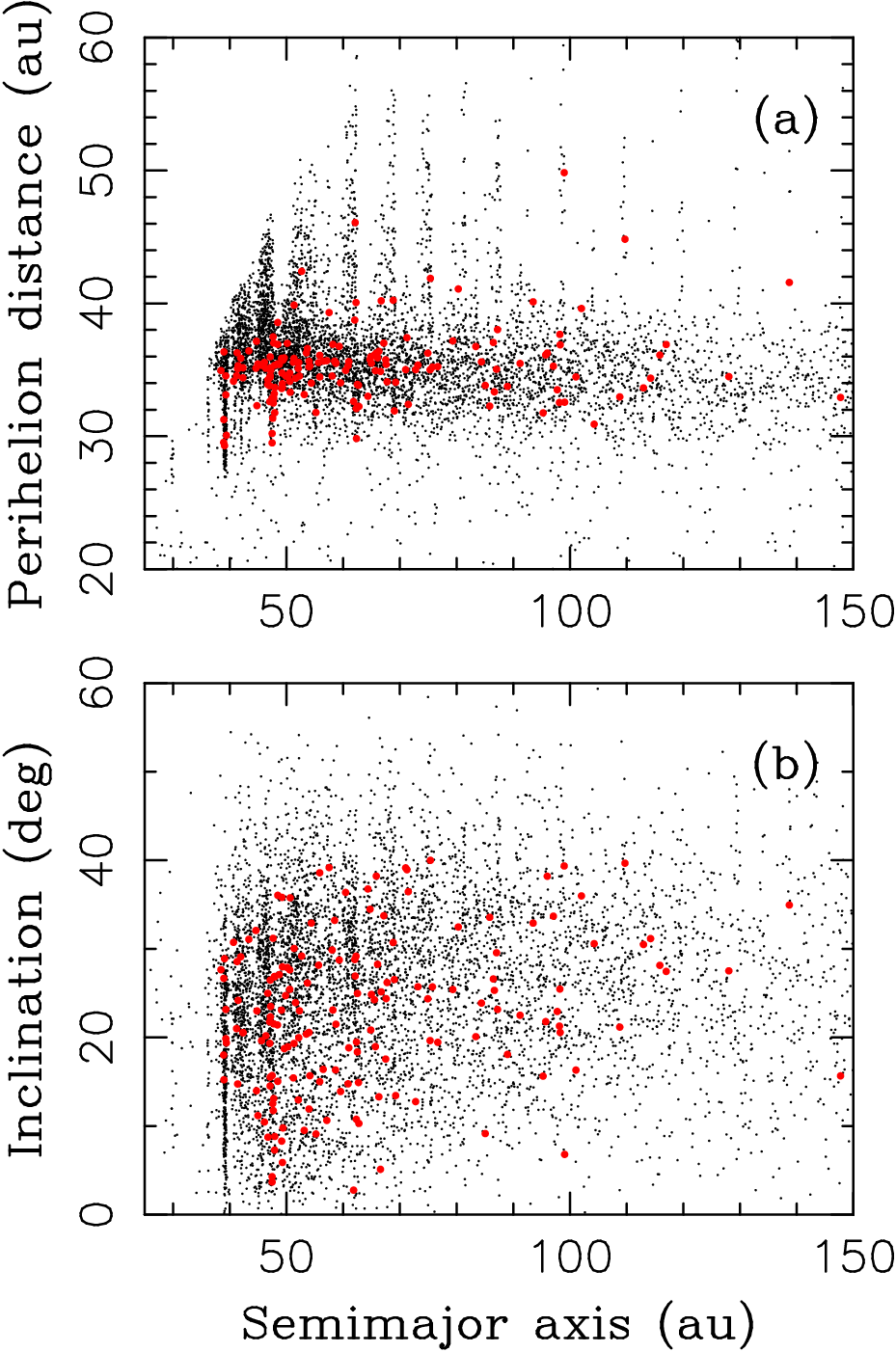}
\end{minipage}
\hspace*{0.0in}
\begin{minipage}{2.2in}
\centering
\includegraphics[width=2.2in]{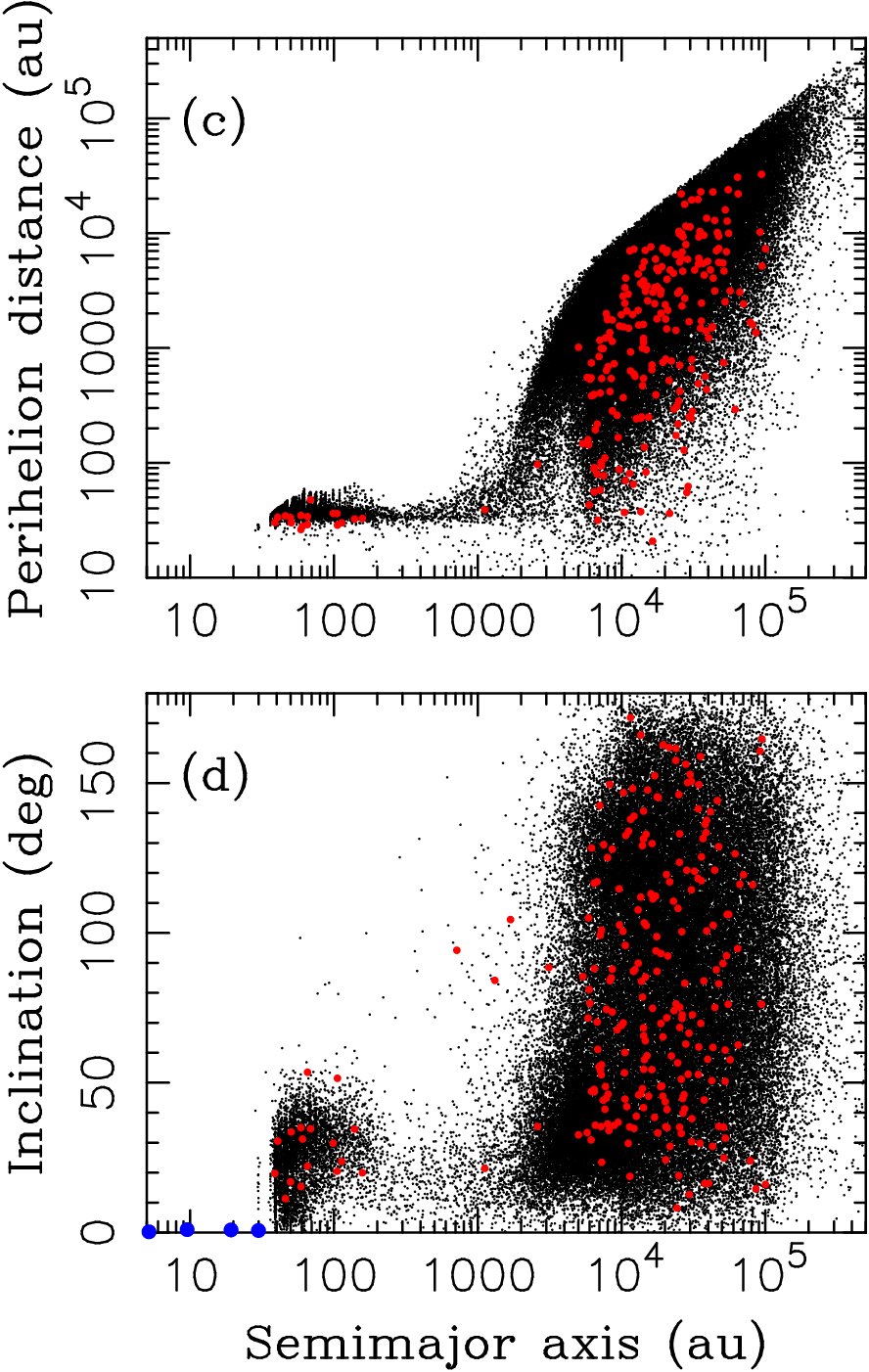}
\end{minipage}
\caption{The orbits of trans-Neptunian bodies that dynamically evolve to become SPCs. The 
source of ECs is shown on the left (panels a and b). The source of HTCs is shown on the right (panels c and d).
ECs were selected using $2<T_{\rm J}<3$, $P<20$~yr and $q<2.5$ au and $N_{\rm p}(2.5)=500$. The 
source orbits of ECs were identified at $t=1.5$ Gyr after $t_0$ (i.e., about 3 Gyr ago), 
and plotted here with red dots. HTCs were selected using $T_{\rm J}<2$, $10<a<20$ au and $q<2$ au and 
$N_{\rm p}(2.5)=3000$. The source orbits of HTCs are plotted at $t=3.5$ Gyr or about 1 Gyr ago.  Background 
orbits are denoted by black dots.}
\label{source}
\end{figure}

The source reservoir of most ECs ($\simeq$75\%) is the scattered disk with $50<a<200$~au (\textbf{Figure \ref{source}}). 
About 20\% of ECs started with $a<50$ au. The classical KB, including various resonant populations below 50 au (about 4\% of ECs 
evolved from the Plutino population), is therefore a relatively 
important source of ECs. Interestingly, $\simeq$3\% of model ECs started in the Oort cloud. 
The orbital evolution of these comets is similar to returning LPCs or HTCs, except 
that they were able to reach orbits with very low orbital periods and low inclinations. The median semimajor 
axis of source EC orbits is $\simeq$60 au. 

HTCs were found to have a nearly isotropic inclination distribution and appear as an 
extension of the population of returning LPCs to shorter orbital periods (N17). The number of large HTCs obtained 
in the model from the Oort cloud agrees well with observations. A great majority 
($\simeq$95\%) of HTCs come from the Oort cloud, and only $\simeq$5\% from the $a<100$ au region (\textbf{Figure 
\ref{source}}). The inclination 
distribution of source orbits in the Oort cloud is slightly anisotropic with the median inclination $\simeq$$70^\circ$. 
This is similar to the median inclination of new HTCs in the N17 model. The inner and outer parts of the Oort cloud
(1,$000<a<20,$000 au inner, $a>20,$000 au outer), contribute in nearly equal proportions to the HTC population. 

The inner scattered disk at $50<a<200$ au should contain $\sim 1.5\times10^{7}$ $D>10$~km bodies, and the Oort cloud 
should contain $\sim 3.8\times10^8$ $D>10$ km comets. These estimates were obtained by calibrating the 
original outer disk from Jupiter Trojans (Section 9) and propagating all populations to the current epoch. 
They are consistent with the number of observed comets that evolve from these reservoirs to the inner Solar System (see 
also Brasser \& Morbidelli 2013). The above estimates can be extrapolated to smaller or larger sizes using the size 
distribution of Jupiter Trojans (\textbf{Figure \ref{sfd}}). 

\section{LATE HEAVY BOMBARDMENT}
The impact cratering record of the Moon and the terrestrial planets provides important clues about the formation
and evolution of the Solar System (Bottke et al. 2017). Especially intriguing is the epoch $\simeq$3.8-3.9~Gyr ago (Ga), 
known as the Late Heavy Bombardment (LHB), when the youngest lunar basins such as Imbrium and Orientale formed (see 
Chapman et al. 2007 for a review).  It has been argued that the impact record 
can be best understood as a `sawtooth' profile (Bottke et al. 2012, Morbidelli et al. 2012b, Marchi et al. 2012), 
a combination of decaying flux from the terrestrial planet accretion leftovers, and a modest increase in the number 
of asteroid/comet impacts (by a factor of 5 or so) produced by a dynamical instability in the outer Solar System (Section 
3). These works assumed that the instability happened late, some $\simeq$3.8-4.2~Ga or $\sim$400-700 Myr after~$t_0$.  

The time of planetary migration/instability cannot be determined from the modeling alone, because it depends on 
unknown details of the Solar System architecture at the time of the gas disk dispersal. The late version of the 
instability/migration was advocated by the proponents of its relation to the LHB (see the references above). It could 
also explain the shock age distribution of various meteorite classes which show a period of enhanced collisional activity 
around the LHB time. The instability could have been delayed if the inner edge of the outer planetesimal disk was
well separated, by a few au, from the outermost planet, such that it took a while for planetesimals to reach the 
planet-crossing orbits (Gomes et al. 2005). It could have been triggered by distant perturbations from the outer 
disk planetesimals (Levison et al. 2011) or by interaction of planets with large amounts of dust evolving inward from
the outer disk by Poynting-Robertson drag (Deienno et al. 2017). 

The early instability, on the other hand, offers several notable advantages. First, the instabilities in dynamical systems 
generally happen early, not late. To trigger a late instability in the outer Solar System, and obtain planetary evolution 
histories that are compatible with the observed structure of the Kuiper belt, the parameters of the outer 
planetesimal disk may need to be fine tuned (Gomes et al. 2005, Deienno et al. 2017). Second, the early version of the 
dynamical instability would relax the terrestrial planet constraint (Agnor \& Lin 2012, Kaib \& Chambers 2016), because 
the secular resonances would sweep through the inner Solar System {\it before} the terrestrial system was in place 
(Section~7). Third, the analyses of Highly Siderophile Elements (HSEs; Kring \& Cohen 2002) and oxygen isotopes (Joy et al. 
2012) do not provide any firm evidence for cometary impactors, while dynamical models indicate that the number 
of cometary impacts during the instability should have dwarfed the asteroid impacts. There is no problem with the cometary impactors 
if the instability happened early, because the early impacts would not be recorded in the geochemical markers.

A recent study of impacts on the terrestrial worlds was reported in Nesvorn\'y, Roig \& Bottke (2017; hereafter NRB17). 
NRB17 developed a dynamical model for the historical flux of large asteroid and comet impactors and discussed how it depends 
on various parameters, including the time and nature of the planetary migration/instability. The orbital evolution of planets 
in different cases was taken from Bottke et al. (2012), NM12 and Roig et al. (2016). They found that the asteroid impact flux 
dropped by 1 to 2 orders of magnitude during the first 1 Gyr and remained relatively unchanged over the last 3 Gyr. The 
early impacts were produced by asteroids whose orbits became excited during the planetary migration/instability, and by 
those originating from the inner extension of the main belt (E-belt; semimajor axis $1.6<a<2.1$ au, Bottke et al. 2012). 
  
\begin{figure}[t]
\centering
\begin{minipage}{2.4in}
\centering
\includegraphics[width=2.3in]{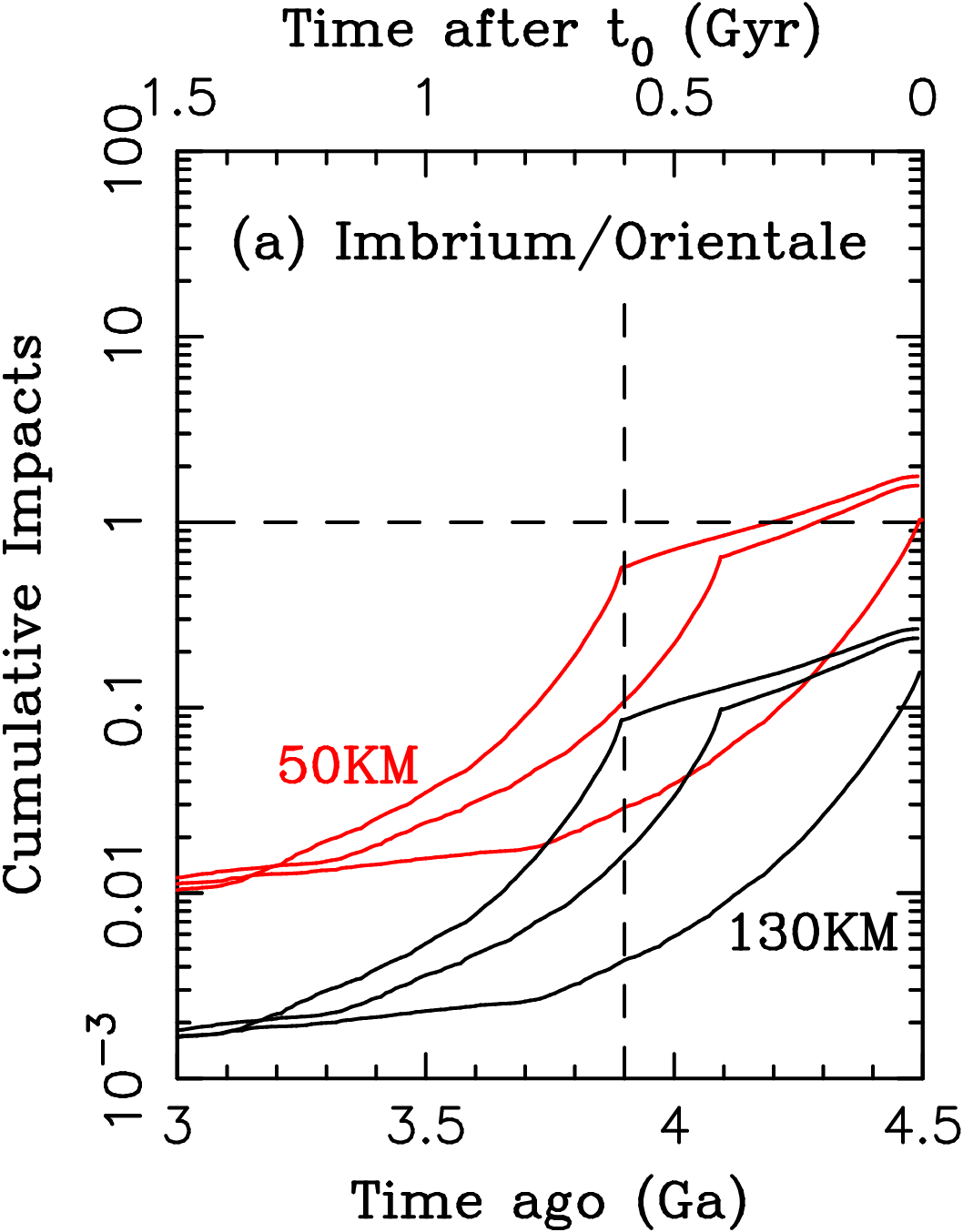}
\end{minipage}
\hspace*{0.1in}
\begin{minipage}{2.17in}
\centering
\includegraphics[width=2.07in]{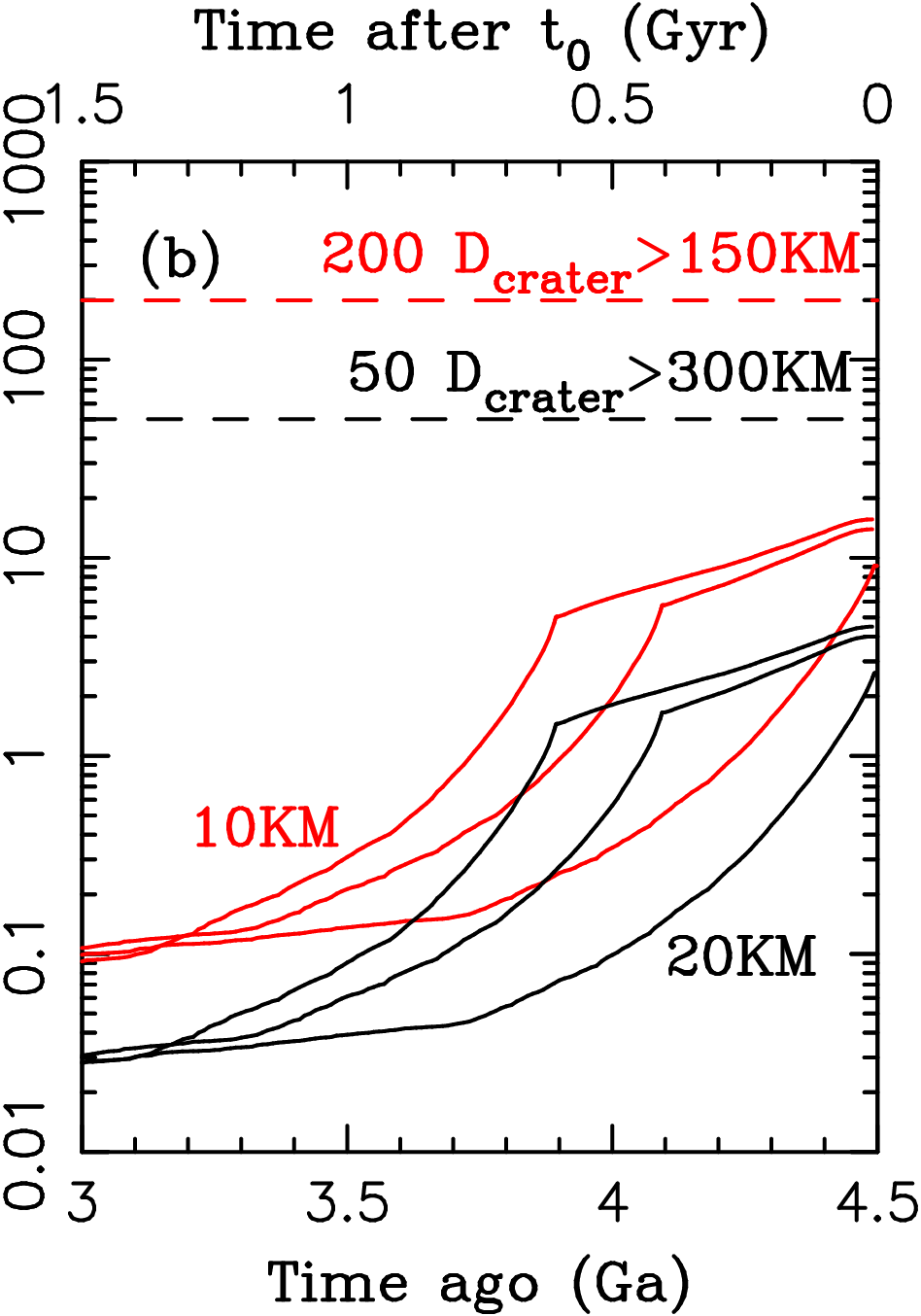}
\end{minipage}
\caption{A comparison of constraints from large lunar craters and basins with the asteroid impact fluxes from NRB17. 
Panel (a) highlights the constraint from the Orientale and Imbrium basins. The family of solid black
curves shows the calibrated impact profile of $D=130$-km (Imbrium) impactors for three different values of the 
instability time ($t_{\rm inst}=3.9$, 4.1 and 4.5 Ga). The red curves show the same for the $D=50$-km (Orientale) 
impactors. Panel (b) reports the results of dynamical modeling relevant for the $D>150$-km lunar craters and $D>300$-km lunar basins, here 
assumed to require $D>10$ km and $D>20$ km asteroid impacts at 23 km s$^{-1}$. The red dashed line in (b) shows the 
number of $D>150$-km lunar craters, obtained by scaling the crater densities from the heavily cratered, ancient 
terrains on the far side of the Moon to the whole lunar surface (Bottke et al. 2017). In total, only $\sim$10-15 
$D>10$-km asteroids should have impacted the Moon, with all these impacts happening during the first 1 Gyr of the Solar 
System history. The Earth and Venus received $\sim$20 times as many, or $\sim$200-300 $D>10$ km asteroid impactors.}
\label{lhb}
\end{figure}

The asteroid impact flux was absolutely calibrated in NRB17. To this end, the number of bodies surviving in the 
asteroid belt at $t=4.5$ Gyr after the start of the integration was compared with the actual number 
of known main belt asteroids of given size (for example, there are $\simeq$8000 asteroids with $D>10$ km; 
Masiero et al. 2014). This resulted in a factor that was applied to compute an absolutely calibrated historical 
record of impacts. The results indicate that asteroids were probably {\it not} responsible for the LHB, independently 
of whether the instability happened early or late, because the calibrated flux is not large enough to explain 
Imbrium/Orientale and a significant share of large lunar craters (\textbf{Figure~\ref{lhb}}). 

Comets (Levison et al. 2001, Gomes et al. 2005) and leftovers of the terrestrial planet accretion (Morbidelli et al. 2001, 2017) 
provided additional, and perhaps dominant source of impacts. The number
of comets in the original outer disk can be absolutely calibrated from Jupiter Trojans.
The cometary impact flux on the terrestrial worlds can then be estimated from dynamical integrations (Gomes et al. 2005, 
NRB17). The results show that the overall impact probabilities of comets on the Earth are $\simeq5\times10^{-7}$
(for each comet in the original disk). With 
$\simeq6\times10^9$ $D>10$ km comets in the original disk (\textbf{Figure~\ref{sfd}}), this implies $\sim$3000 terrestrial 
impacts of $D>10$ km comets. This is $\sim$10-15 times the expected number of asteroid impacts. The Moon
should have received $\sim$1/20 of the terrestrial flux, or $\sim$150 $D>10$ km comet impactors. Since the geochemical 
evidence argues against comets being the main source of the LHB, this may indicate that the instability happened 
early (i.e., before the lunar surface was able to record large impacts).

New terrestrial planet formation models
show that that leftover impact flux probably decayed more slowly (Morbidelli et al. 2017) than 
thought before (Bottke et al. 2007). Unfortunately, the terrestrial planet formation is still not understood well 
enough  
to absolutely calibrate the leftover impact flux. Instead, possible impact histories can be obtained from various 
geochemical constraints. For example, assuming that the HSEs track the total amount of material accreted by the Moon 
since its formation would imply that the LHB was a spike in the bombardment flux (Morbidelli et al. 2012b). If the HSEs 
were depleted from Moon's mantle during magma ocean crystallization, due to iron sulfide exsolution (Rubie et 
al. 2016), however, the LHB can be explained as tail of slowly decaying impact flux of the terrestrial planet 
leftovers (Morbidelli et al. 2017). This would imply that the lunar magma ocean crystallized $\sim$100 Myr after 
the Moon's formation. 
\section{SUMMARY}
Historically, it was thought that the planets formed near their current locations. However, starting with the pioneering 
works of Goldreich and Tremaine (for planet--gas disk interactions) in the late 1970s and Fern\'andez and Ip (for 
planet--planetesimal disk interactions) in the early 1980s, it has become clear that the structure of the outer 
Solar System, at least, most likely changed as the planets grew and migrated. As we have broadened our horizons 
concerning the theory of planet formation, we have significantly increased the size of the parameter space that we 
need to explore. As a result, we have gotten to the point where we need to use any available constraint on the problem.

In sections 6-13, we discussed various constraints on the dynamical evolution of the early Solar System. 

Section 6: The obliquities of the giant planets constrain the behavior of the $s_7$ and $s_8$ frequencies that govern the
orbital plane precession of Uranus and Neptune. The frequency $s_8$ is required to evolve fast initially and very slowly 
toward the end of planetary migration. This happens in most migration models as Neptune's migration tends to slow down 
over time. At late stages, the effective e-folding migration timescale must have been $100<\tau<200$ Myr for Saturn's
obliquity to be excited by capture in the spin-orbit resonance with $s_8$.

Section 7: The terrestrial planets are a very sensitive probe of the behavior of the $g_5$ frequency that controls the 
precession of Jupiter's orbit. If the instability happened late, when the terrestrial planet system was already in place, 
the resonances of $g_1$ (Mercury) and $g_2$ (Venus) with $g_5$ must be avoided, because they would lead to an excessive 
orbital excitation of the terrestrial planets. This can be best accomplished in the jumping-Jupiter model where 
the $g_5$ frequency has a discontinuity as a result of planetary encounters.
 
Section 8: The asteroid belt constraints suggest that the $g_6$ (precession of Saturn's orbit) and $s_6$ (precession Saturn's 
orbital plane) frequencies changed fast such that the $s=s_6$ and $g=g_6$ resonances spent much less than 1 Myr in 
the 2-3 au region. This happens in the jumping-Jupiter model, where the period ratio $P_6/P_5$ changes 
from $<$2 to $>$2.3 in $\ll$1 Myr, and the resonances only briefly overlap with the inner part of the
asteroid belt ($a<2.5$ au). The asteroid constraint applies independently of whether the instability happened 
early or late. 

Section 9: From modeling the planetary migration/instability, NM12 estimated that the mass of the outer disk of planetesimals 
(23-30 au) was $M_{\rm disk}\simeq15$-20 $M_\oplus$. An independent estimate of $M_{\rm disk}$ can be obtained from the 
capture probability and present population of Jupiter Trojans. This estimate also gives $M_{\rm disk}\simeq15$-20 
$M_\oplus$, providing support for the NM12 model.  
 
Section 10: The moons of the giant planets represent important constraints on planetary encounters during the instability. 
The orbital inclinations of Uranus's regular moons suggest that Uranus probably did not have close encounters with 
Jupiter/Saturn. The Galilean moons at Jupiter imply that there were no encounters between Jupiter and an ice giant with 
mutual distance $d<0.02$ au. The similar populations of irregular satellites at different planets suggest that all 
giant planets had at least several distant encounters with another planet.
 
Section 11: The Kuiper belt constraints imply that Neptune's migration was slow ($\tau \geq 10$ Myr), long range ($a_{8,0}\leq25$ 
au) and grainy (1000-4000 Pluto-class objects in the original outer disk). The migration timescale implies that the
outer disk at 23-30 au had mass $M_{\rm disk}\simeq15$-20 $M_\oplus$, which is consistent with several independent 
estimates discussed above. Encounters of Neptune with the fifth planet, which produce a discontinuity during Neptune's 
migration, can explain the Kuiper belt kernel.   

Section 12: Modeling implies that the inner scattered disk at $50<a<200$ au should contain $\sim 1.5\times10^{7}$ $D>10$~km 
bodies. The Oort cloud should contain $\sim 3.8\times10^8$ $D>10$ km comets. The number of large JFCs and HTCs obtained
from these reservoirs in the model is consistent with the number of observed comets. To fit various constraints, the 
physical lifetime of comets must be a strong function of comet size. 

Section 13: The absolute calibration of the impact flux indicates that asteroids were probably not responsible for the LHB,
independently of whether the instability happened early or late, because the calibrated flux is not large enough to 
explain Imbrium/Orientale and a significant proportion of large lunar craters. Therefore, attempts to delay the 
instability may not be useful. Comets and leftovers of the terrestrial planet accretion probably provided a dominant 
source of impacts on the terrestrial worlds during early epochs. 

Although significant uncertainties remain, the Solar System constraints discussed above allow us piece together the 
following sequence of events in the early Solar System history. The outer planets emerged from the protoplanetary 
gas disk in a compact resonant configuration with Jupiter and Saturn in the 3:2 (or 2:1) resonance and Neptune at 
$\simeq20$-25~au. The third ice giant planet with the mass similar to Uranus or Neptune was located near $\simeq10$ au 
(between the original orbits of Saturn and Uranus). The outer planetesimal disk extended from just outside of Neptune's 
orbit to $\simeq$30 au, and had a low mass extension reaching to at least $\simeq$47 au. Several independent estimates 
suggest $M_{\rm disk}\simeq15$-20 $M_\oplus$ and the SFD of the disk planetesimals similar to that of present-day Jupiter 
Trojans. 

Within a few tens of Myr, Neptune slowly migrated into the planetesimal disk, scattering planetesimals around, and taking 
$>$10 Myr to reach its current orbit near 30 au. The other planets migrated as well. During the planetary migration, 
probably when Neptune reached $\simeq$27.7 au, a dynamical instability occurred with the fifth planet having encounters 
with all other outer planets. It was subsequently ejected from the Solar System by Jupiter. As a result, Jupiter's semimajor 
axis changed by a fraction of an au and Jupiter's proper eccentricity mode was excited to its current value. The inner 
part of the asteroid belt was destabilized by resonances with Jupiter and contributed to the early impacts in the inner 
Solar System. If the terrestrial planet formation was not completed at this point, these early impacts would not be 
recorded on surfaces of the terrestrial worlds. Mercury's orbit may have been excited during the instability. 

In the end, the outer planetesimal disk was completely dispersed by planets and parts of it were implanted into different 
populations, including the asteroid belt (estimated implantation probability $5\times10^{-6}$), Jupiter Trojans ($5\times10^{-7}$), 
Jupiter irregular satellites ($2.5\times10^{-8}$), Kuiper belt ($10^{-3}$), scattered disk ($3\times10^{-3}$), and the 
Oort cloud ($5\times10^{-2}$). The size and orbital distributions of bodies in these populations, either observed 
or inferred indirectly, are consistent with the planetary migration/instability model discussed above. In fact, as we 
argued throughout this text, the populations of small bodies in the Solar System represent fundamental constraints on the 
early dynamical evolution of the Solar System. The model of planetary migration/instability outlined above was developed 
in an attempt to satisfy them all.

\section{DISCLOSURE STATEMENT}
The authors are not aware of any affiliations, memberships, funding, or financial holdings that might be perceived as affecting 
the objectivity of this review. 

\section{ACKNOWLEDGMENTS}
This work was was sponsored by the NASA Emerging Worlds program. Luke Dones, Alessandro Morbidelli and David Vokrouhlick\'y 
supplied many helpful comments to the manuscript. I dedicate this text to my father Jan. 


\end{document}